\documentclass[12pt,a4paper]{article}

\usepackage{amsfonts}
\usepackage{amsxtra,amssymb,amscd,amsthm,footnote}

\begin{document}
\author{Jean-Marc OURY\footnote{Jean-Marc OURY, Ingenieur au corps des mines, Mod\'{e}lisation, Mesures et Applications (MOMA), Paris, is the
inventor of the Absolute Relativity Theory. Email: art@momagroup.com}, Bruno HEINTZ\footnote{Bruno HEINTZ, Ingenieur au corps des mines, Mod\'{e}lisation, Mesures et Applications (MOMA), Paris, collaborated to the development of the Absolute Relativity Theory since 1996 by contributing
to its mathematical foundations and by checking, improving and completing
the methods and computations required for the Mass Quantification Theory. Email: art@momagroup.com}}

\title{The Absolute Relativity Theory}
\date{August 2009}
\maketitle

\begin{abstract}
The purpose of this paper is to give a first presentation of a new approach
of physics, that we propose to refer to as the Absolute Relativity Theory.

This approach is founded on the refutation of the old idea that our universe
can be seen as a space-time, whatever structure it is equipped with, that
contains or supports ``observers'' and ``observables''.

Instead, the theory begins by exploring what should be, from an algebraic
point of view, a consistent theory able to represent the ``observation''
processes and, in some sense, as complete as possible. Expressed in the
general framework of categories, those conditions lead to take as a basis of
the theory the category $\aleph$ of all the finite dimensional
representations of complex quasi-Hopf algebras, and to explore its morphisms
that appear as natural transformations. In this framework, the relation
between observables and observers is naturally represented by the usual Hom bifunctor that associates with any couple of objects the set of
arrows between them. Its contravariant part is called the point of view
functor since its describes the arrows that arrive to this object.

Two principles are then stated in order to link the theory with physics,
that we propose to refer to as the Absolute Relativity Principle and the
Absolute Equivalence Principle. They express that any natural transformation
in $\aleph $ corresponds to a true physical phenomenon, and reciprocally.
Taken together, those principles lead to think that physics should be the
description of the structures in $\aleph $ as seen through the point of view
functor.

This is the aim of our first section that process in twelve successive steps
that can be summarized as follows.

Since Hopf algebras can be classified according to their primitive part, the
well known classification of real Lie algebras leads to determine a set of
compatible ones, the representations of which could correspond to usual
particles. The characterization of each type of these real Lie algebras through its Cartan
subalgebra leads to associate with each type a specific space-time that
is linked to the others by the usual embedding of algebras.

Two mathematical facts play then a key role. The first one is a specific
property of the endomorphisms of $\mathbf{sl}_{n}$-algebras, seen as acting
on those Cartan subalgebras that define different space-time, that have an order two
specific symmetric representation that can be identified with the Lagrangians of
physics. The second one is the fact that the point of view functor, being
contravariant, induces the change to dual Hopf algebras that have infinite
dimensional polynomial-type representations. Any object is represented in
those algebras as inducing a creation/annihilation operator that generates
the passing of his time. This mathematical fact corresponds globally to the
so-called ``second quantization'' introduced by Quantum Field Theory. The
use of Yangians representations gives a useful tool for representing this
operator as its infinite dimensional generator $J$.

The capacity to algebraically represent Lagrangians and the ``passing of the
time'' leads naturally to define physical observers as objects in $\aleph $
together with a well defined Lagrangian and passing of time. The
identification with usual physics can now be made. 

Euler-Lagrange equations first appear as the expression in our quantized
context of the monodromy of the Knizhnik-Zamolodchikov equations. The well-known
relations between Lie algebra together with the representation of the passing of
time gives a computation process, that should permit to calculate from only one measured data, the fine
structure constant $\alpha$, the characteristics of the particles (like their mass) as
identified by contemporary physics.

The theory also predict the existence of non identified (as of today) other particles
that could give new interpretations of the ``dark matter'' and the ``dark
energy''. All those computations may be seen as a new theory that explains
why and how the matter does appear in a well defined quantized way : We
propose to call this new branch of physics the Mass Quantification Theory.

The first section finally mentions that if one wants to represent space-time
from the point of view of a physical observer as a Lorentzian manifold, the identification of local monodromies on this manifold with the above KZ-monodromy applied to our algebraic Lagrangians gives exactly the Einstein
equation, but now its right-hand side has also found a mathematical interpretation.
Since it comes from the quantized part of physics, this interpretation could
be seen as what introduces General Relativity in the realm of quantum physics, i.e. as the unification of the two.

The second section of this paper is dedicated to a more mathematical
presentation of the foundations of the Absolute Relativity Theory.

The third section contains the first computations that can be obtained from
the Mass Quantification Theory. The almost perfect concordance of
the results obtained by pure calculations with the best experimental values can be seen as promising
for future progresses.

\end{abstract}

\begin{center}
$\mathbf{Acknowledgement}$
\bigskip
\end{center}
The authors are much grateful to Professor Sergiu Klainermann (Princeton University, Department of Mathematics) for many exchanges about the work presented in this paper. At various stage of the development of the theory, Jean-Marc Oury presented the evolutions of his ideas to Professor Sergiu Klainermann, and benefited from his questions and as well as his pointing out aspects of the theory that required further clarifications.

\pagebreak

\begin{center}
\emph{``This does not seem to be in accordance with a continuum
theory, and must lead to an attempt to find a purely algebraic theory for
the description of reality. But nobody knows how to obtain the basis of such
a theory.''}\footnote{Albert Einstein, The Meaning of Relativity,
sixth edition, Routledge Classics, London, 2003, p.170. }
\end{center}

\section{Introduction to the Absolute Relativity Theory}

\subsection{The status of space-time in physics}

By refuting the old Aristotelian distinction between to be ``at rest'' or
``in motion'', the classical relativity principle asserted that ``motion''
can only be defined relatively to Galilean ``observers'', which presupposes
the existence of a Newtonian ``space'' that defines those observers and
contains observable ``objects''.

Three centuries later, Einstein's special relativity theory replaced
Galilean observers by Lorentzian ones and extended space to
space-time, without altering the universal character of the latter.

Only the General Relativity theory began to inverse the factors by stating that
space-time does not contain the observable ``objects'' but \emph{is
defined by them in an interactive process encoded by Einstein's equation
that links Ricci curvature tensors of space-time seen as a Lorentzian
manifold with the distribution of energy (or masses) induced by those
objects. }But the price paid for this reversal was quite high since the use
of Ricci curvature tensors prevented the theory from describing internally
asymmetric phenomena. So, until recently, general
relativity theory has remained an isolated branch of contemporary physics,
mainly dedicated to the study of gravitational phenomena.

The second branch of contemporary physics, born with Schr\"{o}dinger,
Heis\-enberg and Dirac, kept asserting the intrinsic ``existence'' of a
space-time, thought as ``the referential of the laboratory of the observer''
on which Hilbert spaces of $\mathbb{C}$-valued functions are defined. Only
later, with the emergence of quantum electrodynamics (QED) and quantum
fields theory (QFT), systematic efforts were made in order to reverse the
factors by assuming for instance that ``observable'' could also be defined
as elements of abstract algebras, and that ``states of nature'' may be seen
as $\mathbb{R}$-valued operators on such algebraically predefined
``observable''.

Finally, the status of space-time in string theories is even more
astonishing since its existence is a prerequisite for each of them, but its
dimension and folded structure differ from one theory to the other. So, despite
many crucial discussions about space-time, string theories do not really
challenge its intrinsic existence either\footnote{{\small In the same way,
the recent loop quantum gravity theory (LQG) assigns a specific ``loop''
structure to  space-time, but does not challenge either its a priori
existence. Its first successes demonstrate how crucial for physics is the
problem of space-time.}}.

\emph{Thus, the modern and contemporary idea of the intrinsic existence of
some (generalized) space-time could well be the Gordian knot that has been
blocking the path to the unification of physics for one century.}

Since this idea amounts to representing the space-time perceived by the set
of observers we belong to as universal, we propose to refer it as \textbf{
the egocentric postulate}.

Unfortunately, since this egocentric postulate always stands at the very
beginning of physical theories, refuting it immediately leads to the
question of the foundations of the construction we intend to build.

\subsection{Observers, observables, and physical phenomena}

The theory presented here answers this question \emph{by noticing that the
first step in the construction of a physical theory should be to define a
representation of the observation process itself, which means a definition
of observers and observables together with a description of their relations.}

The algebraic approach suggests \emph{to treat observers and observables
on the same footing by seeing them as objects of an appropriate category},
that we propose to call \textbf{the preuniverse} $\aleph$, and defining
the \textbf{point of view of an object} $a$ \textbf{(observer) on an other
object} $b$ \textbf{(observable)} \emph{as the usual set of arrows from b to a:}
 $\hom(b,a)$. \emph{Any object will thus
be both an observer and an observable}.

We will call the usual contravariant functor $\hom (.,a)$ \textbf{the
point of view functor of }$a$, and any functor $F$ equivalent in $\aleph $
to $\hom (.,a)$ will be said to be \textbf{represented by} $a$. The
consequences of the contravariance of the point of view function cannot be overstated since any
object, as an observer, belongs to the preuniverse $\aleph $, but, as an
observable, it belongs to the opposite category $\aleph ^{opp}$ defined by
``reversing the arrows''. In other words, from the very first algebraic
foundations of the theory, physical objects have to be described with a
``double nature'', first as observers in $\aleph $, second, as they
are seen from other objects, namely objects in $\aleph ^{opp}$.

By considering then some internal consistency conditions (mainly requiring
appropriate ``diagrams of arrows'' to be commutative) and completeness
conditions (mainly requiring the preuniverse to have ``enough'' objects to
represent ``composed'' objects and point of view functors), we will precise
the structure of the preuniverse $\aleph $ that will appear as a well
defined category of\textbf{\ representation functors }$\rho _{A,V}$ (or $A$
\textbf{-bimodules}) from various complex Hopf algebras $A$ to finite
dimensional bimodules $V$.

$\aleph ^{opp}$ is then the dual category $\aleph ^{*}$, that is the
category of (non necessary finite dimensional) bimodules over various dual
algebra $A^{*}$ (in the sense of the restricted Hopf duality). Both algebras $A$
and $A^{*}$ fundamentally differ : for instance, non trivial $A$-modules
we will consider have a greater than one finite dimension, while $A^{*}$
-modules are either one- or infinite- dimensional. Therefore, the above
``double nature'' of objects  will
have tremendous consequences\footnote{{\small Representations of Hopf
algebras are bimodules that reflect both the algebra and coalgebra
structures of the Hopf algebra (including their compatibility).
Nevertheless, the usual way to deal with coalgebraic structures is to see
them from the dual point of view, i.e. as an algebraic structure on the dual
Hopf algebra. We will proceed so, and correlatively use the word ``module''
on one of those algebras, instead of speaking of ``bimodule'' and precising
which of the algebraic or coalgebraic structures we refer to.}}.

\emph{\ Hopf algebras are generally not
cocommutative. This means that their dual cannot be seen as a space of
functions, and thus that our categorial approach leads directly to so-called
``quantized'' (or ``deformed'') situations.}

We will get in fact\emph{\ two type of quantizations}, the first one associated
with the non cocommutativity of $A$ will globally correspond to the one of
quantum mechanics, the second, more subtle, associated with the non
cocommutativity of $A^{*}$, will generate creation/annihilation operators
that correspond to the ``second quantization'' introduced by the QFT.

Let us emphasize that \emph{those quantizations do not appear in
cocommutative algebras that therefore cannot be seen as basic cases, but
as degenerate ones, which explains the devastating consequences of the
egocentric postulate}.

By working indeed on operators on an Hilbert space of functions, QFT
is compelled to use ``perturbational methods'' in order to generate
quantized situations, instead of representing them directly, as we will do
below, \emph{by noticing that such quantized situations intrinsically induce the
existence of creation/annihilation operators that, in some mathematical
sense we describe precisely, produce at each present instant the
immediate future of particles and associated space-time.}

Finally,\emph{\ a physical theory has to be able to be confronted with
experimental results. }This implies that the theory includes a
representation of what the universe we perceive is, which we propose to
refer to as made of \textbf{true physical objects} and \textbf{true physical
phenomena}, and that it describes the impact of the perception process from
the point of view of the observers we are. Considering our purpose, we will
not view them as embedded in any given space-time, nor as having any a
priori given proper time.

To this aim, \emph{we will postulate that all the true physical objects
are representation functors }$\rho _{A,V}$ \emph{for appropriate algebras }
$A$\emph{\ and modules }$V$ \emph{that we will precise.}

We will then use two (and only two) principles to characterize true physical
phenomena, namely\emph{\ any natural transformation between representation
functors induces a true physical phenomenon, and, reciprocally, any physical
phenomenon can be seen in the theory as induced by a natural transformation.}

\emph{Since those principles will appear as the ultimate extensions of
Einstein's relativity and equivalence principles}, we propose to call this
new theory \textbf{the Absolute Relativity Theory (ART)}, and to refer to
these two principles as the \textbf{ART principles}.

Finally, as predicted by contemporary theories, we will represent ourselves
as made of objects we will identify with electrons, and nucleons (
made of quarks linked by strong interaction, which is not a trivial fact
since it implies, among others, that \emph{there are objects and
interactions that arise from algebras bigger than the ones we belong to : we
necessarily perceive them not as they are, but only through ``ghost
effects'' as representations of the particles we are made of}). The weak
interaction will appear as belonging to this category, which explains why
its bosons do \emph{appear} to us as massive, and induces the existence of
an energy that we do not perceive as it is, and that therefore could be part
of so-called ``dark energy''.

\emph{In the ART framework, the way the observables do appear
to observers does not only depend on their relative positions and motions,
but also on the type of each observer}.

Here is the way we will follow.

\subsection{A short description of the Absolute Relativity Theory}

The next subsection of this paper gives a tentative twelve steps presentation of
the main foundations and results of ART.

We will first precise the definition and characteristics of the above
preuniverse $\aleph $, and see how a first classification of its ``objects''
and ``morphisms'' leads to a first classification of particles and
interactions as associated with some specific \emph{real} form of simple
finite dimensional complex algebras. Ultimately, ``our perceived'' space-time itself will
arise in a very natural way from those algebraic foundations.

\emph{A specific emphasis has to be made on the algebraic meaning of the well
known Lagrangians, and on the origin of the ``passing of time'' phenomenon}.

The first will appear as the consequence of a very specific property of the
representations of the algebras $\mathbf{sl}_{n}\otimes \mathbf{sl}_{n}$ (
with $n>2$) that admit, as a representation of $\mathbf{sl}_{n}$,\emph{\ a
trivial component that we will call \textbf{the free part of the Lagrangian }
and a non trivial symmetric component that we will refer to as\textbf{\ its
interactive part}.}

\emph{The free part defines an invariant bilinear symmetric form on the
dual of the representation space on $sl_{n}^{\ast}$. Since we will
apply }$\mathbf{sl}_{n}$\emph{\ to the Cartan subalgebras }$\mathbf{h}$
\emph{\ of the Lie algebras }$\mathbf{g}$\emph{\ (of rank }$n$\emph{)
we will work on, this free part will correspond to a (complex) scaling to be
applied to the standard restriction to }$\mathbf{h}$\emph{\ of the dual of 
}$\mathbf{g}$\emph{\ Killing form. It will appear as directly connected to
mass and charge of particles.}

\emph{The interactive part is more subtle, as defined by the following
projection operator }$\pi $\emph{\ (viewing elements of }$\mathbf{sl}_{n}$
\emph{\ as matrices) :}

\begin{center}
$\pi (x\otimes y)=xy+yx-\frac{2}{n}trace(xy).Id$.
\end{center}

\emph{It associates in a quadratic way with any element of }$\mathbf{sl}
_{n}$\emph{\ a bilinear symmetric form on }$\mathbf{sl}_{n}^{*}$. \emph{
Identifying }$\mathbf{sl}_{n}$\emph{\ with }$\mathbf{sl}_{n}^{*}$\emph{\
through its own Killing form, we get so a bilinear symmetric form on }$
\mathbf{sl}_{n}\otimes \mathbf{sl}_{n}$ \emph{itself. Identifying it,
again canonically through the Killing form, with }$\mathbf{sl}_{n}\otimes 
\mathbf{sl}_{n}^{*}$\emph{, we get a bilinear symmetric form defined on
endomorphisms of }$\mathbf{sl}_{n}$\emph{. This property is easily
transferred to }$\mathbf{su}_{n}$-\emph{algebras when working on maximal
tori of real compact Lie algebras.}

\emph{Taken together, the free and interactive parts of Lagrangians define
therefore a bilinear symmetric form on the endomorphisms of }$\mathbf{gl}_{n}
$\emph{. The importance of Lagrangians comes from the fact that scalings
and outer automorphisms of a Lie algebra }$\mathbf{g}$\emph{\ have a non
trivial action on the endomorphisms of its Cartan subalgebra, while all the
inner automorphisms by definition do not have any action.}

\emph{On another hand, the ``passing of time'' will appear as directly
coming from the contravariance of the point of view functor that induces for
any simple Lie algebra }$\mathbf{g}$\emph{\ the passing from the
Drinfeld-Jimbo algebra }$U_{h}(\mathbf{g})$\emph{\ to its dual, the
quantized algebra }$F_{h}(\mathbf{G})$ t\emph{hat has a family of infinite
dimensional representations in the formal algebra of one variable
polynomials. This family is in fact induced through the principal embedding
of }$\mathbf{sl}_{2}$\emph{\ into }$\mathbf{g}$\emph{\ by only one case that
arises in }$F_{h}(\mathbf{SL}_{2})$\emph{\ by applying the Weyl symmetry
to }$U_{h}(\mathbf{sl}_{2})$\emph{\ before taking the dual.}

By considering the point of view of an object on itself and thus going from $
U_{h}(\mathbf{g})$ to $F_{h}(\mathbf{G})$ due to the contravariance of the
point of view functor, we have first to notice that from the uniqueness of
principal embeddings of $\mathbf{sl}_{2}$ in the Weyl chamber of any complex
algebra $\mathbf{g}$, we get a family, indexed by the maximal torus and the Weyl group, of
surjection from $F_{h}(\mathbf{SL}_{2})$\ to $F_{h}(\mathbf{G})$. Focusing
thus on the situation in $F_{h}(\mathbf{SL}_{2})$, we see the apparition of
a new type of\emph{\ creation/annihilation operators that recall those of
the Quantum Field Theory (QFT), but applied to the considered object itself,
creating so a new instant after each instant, inducing therefore the
``passing of time''of this object}. Time itself (and thus space-time) so
appears in the ART as defined from objects in a quantized way : in that
sense the ``second quantization'' of QFT should be thought as the first one.

Furthermore, since those representations are indexed by the maximal torus of
the compact form $\mathbf{g}^{c}$ of $\mathbf{g}$, and by elements of the
Weyl group, we will pull back the main characteristics of the real algebra $%
\mathbf{g}$ on the maximal torus $T$ of $\mathbf{g}^{c}$ by specifying its
dimension, its Weyl invariance group, the induced action of the outer
automorphisms of $\mathbf{g}$, and the involution that defines the
considered real form\footnote{{\small All those elements come from their
equivalent in the complex Cartan subalgebra, and we will use systematically
the equivalence between working on complex Lie algebras or on their compact
real form.}}. Furthermore, the action of the above creation/annihilation
operator that generates the passing of time can also be represented on the dual
of the Cartan subalgebra (or maximal torus) of $\mathbf{sl}_{2}$ (or $
\mathbf{su}_{2}$) as inducing the increase/decrease of two units on the line diagram that
corresponds to making the tensor product by the adjoint representation. 

\emph{This leads to carry on with the story by working on $\mathbf{su}_{n}
$-type algebras seen as acting on the maximal torus of $\mathbf{g}
^{c}$}.

\emph{Now, }$\mathbf{su}_{n}$\emph{-type algebras are precisely those on
which Lagrangians can be defined}. They furthermore correspond to a common
framework where everything concerning the different algebras $\mathbf{g}$
can be encoded, and, since Lagrangian as above defined, have the aforesaid
wonderful property to be invariant under $\mathbf{su}_{n}$ adjoint action,
they also give rise to something that will be independent of the chosen
point of view, and that will correlatively be thought as independently
existing : \emph{here is the explanation of the feeling we have that
``things'' do actually exist, ...and the origin of the egocentric postulate}.

The above connection between creation/annihilation operators and polynomial
algebra then leads to work on \textbf{Yangians algebras} defined from those $%
\mathbf{su}_{n}$ algebras. \emph{In this framework, the above creation
operator corresponds to the endomorphism }$J$\emph{\ of }$\mathbf{su}_{n}$%
\emph{\ used to define Yangians}\footnote{{\small See for instance
[V.C.-A.P.], pp.374-391 for all notations and results concerning Yangians.}}%
. \emph{Successive applications of this operator become successive
applications of }$J$\emph{\footnote{{\small In order to represent also the
annihilation operator, one should work on affine algebras instead of
Yangians. Since both are closely related, we will not explore this more
general way in this first presentation of ART.}}}.

\emph{A key point is that, by identifying $\mathbf{su}_{n}$ with 
$\mathbf{su}_{n}^{*}$ through the Killing form, $J$ itself may be seen as an element of $\mathbf{su}_{n}\otimes \mathbf{su}_{n} $and has thus a Lagrangian projection, which defines a relative
scaling of $\mathbf{su}_{n}$ that is very small (we will compute
it as being of order of $10^{-37}$), but exists. We will show that this
fact makes the gravitation appearing as a (quantized) consequence of the passing
of time that is also quantized, as we just have seen.}

\emph{More generally, one can study the evolution of the Lagrangian with
successive applications of the creation/annihilation operator : this
corresponds to the passing from a weight }$N$\emph{\ to the }$N+2$\emph{%
\ one, and therefore, as transposed in a smooth context, to
twice the left-hand side of QFT Euler-Lagrange equation, namely }$\frac{%
\partial \mathcal{L}}{\partial \phi }$.

\emph{But there is a consistent way to go from one representation of }$%
U_{h}(\mathbf{sl}_{2})$\emph{\ to another one that corresponds to the
above passing of time}$,$\emph{\ namely to use Knizhnik-Zamolodchikov
equations (KZ-equations) in order to define the parallel transport of
Lagrangians and get a monodromy above the fixed point of an appropriate
configuration space}.\ Since in the $\mathbf{sl}_{2}$ case, Lagrangians
reduce to their scalar part (the so-called ``free part'') defined by a projector
analog to the QFT quantity $\frac{\partial }{\partial [D_{i}\phi ]}$, the
monodromy of the KZ-parallel transport of this $2$-tensor has precisely to
be equal to the above right-hand side of the QFT Euler-Lagrange equations. Computing this monodromy from the
canonical invariant $2$-tensor associated with $\mathbf{g}$ gives therefore
the true value of the variation of Lagrangian between the two successive
positions, that corresponds to twice the right-hand side of Euler-Lagrange equations. 
\emph{Equalizing both sides and considering any possible direction in
space-time gives Euler-Lagrange equations}.

\emph{Those equations therefore appear as describing in the usual
continuous and smooth context of differential geometry, the above well
defined discrete algebraic creation process that generates the passing of time.
Euler-Lagrange equations can therefore no more be seen as some mathematical
expression of the old philosophical Maupertuis's principle of least action}.

Furthermore, KZ parallel transport implies successive applications of the
flip and antipode operators. Now, two successive applications of the
antipode in a Yangians algebra are equivalent to a Yangian-translation of $
\frac{c}{2}$, with $c$ the Casimir coefficient of the adjoint representation
: in the $\mathbf{sl}_{2}$-case, this translation is by $\frac{8}{2}=4$ ,
which confirms that KZ-parallel transport has to be made with the above two
by two steps of the adjoint representation. KZ-parallel transport therefore
defines a motion on the so called ``evaluation representations'' of the
algebra $Y(\mathbf{su}_{n})$ that sends it back to $U(\mathbf{su}_{n})$ (in
a consistent way with the inclusion map $U(\mathbf{su}_{n})\rightarrow Y(
\mathbf{su}_{n})$).

\emph{These algebraic views can be used to come back to
Einstein's equation and the unification of physics}. Indeed, if it is 
\emph{a priori }postulated that space-time \emph{is }a Lorentzian
manifold, its local monodromy algebra is encoded in the space of Riemann
curvature tensors of which the symmetric part is the Ricci curvature $2$
-tensor. This tensor may thus be seen as the equivalent of the above
right-hand term of the Euler-Lagrange equation that arises, in our context, from
KZ monodromy.

\emph{Writing the equality of both monodromies gives exactly the Einstein
equation. Instead of being seen as a physical principle, Einstein equation
therefore appears as a mathematical consequence of the way the ``passing of the
time'' do arise. As a by-product we get simultaneously an answer to the
question of the origin of the ``arrow of time'' in General Relativity (GR) :
namely, only the ``past part'' of GR Lorentzian manifold that represents the
universe does exist from the point of view of an observer, and this is
precisely the ``passing of time'' of this observer that contributes locally
to extend it.}

\emph{All this leads to the theoretical possibility to compute the Newton constant $G$ from the value of the fine structure constant }$\alpha 
$\emph{\ and the consistency of the theoretical result with the measured value
of this constant could be seen as the symbol of Einstein's come back in the
quantized universe, giving to General Relativity Theory its right place in
quantized physics.}

It makes thus sense to explore the consequences of the above interpretation
of the weak interaction as a ``ghost effect'' from a higher order four
dimensional today unknown interaction. \emph{The introduction of the
corresponding Lagrangian gives a new interactive term that could correspond
to the mysterious ``dark energy''}, but we did not try to evaluate this term
by introducing it in Einstein's equation, nor to test this conjecture. We
also did not try to explore systematically the cosmological implications of
ART.

T\emph{he last step is to use ART to explain the nomenclature and compute
the characteristics of particles and interactions. This is the purpose of a
new branch of physics that we propose to refer as the Mass Quantification
Theory}. With the exception of the above new interpretation of weak
interaction as a ``ghost effect'' from an higher order one, it should have
as an objective to compute and precise the nomenclature and characteristics
of the particles as recognized by contemporary physics. \emph{MQT also
leads to conjecture the existence of new types of particles that appear as
good candidates to be constituents of the ``dark matter''}, but we did not
try to test this conjecture.

Although the MQT theoretically leads to describe and compute any phenomenon
it predicts, only some computations and predictions are introduced in this
first paper.

\emph{We summarize in the next subsection, as twelve successive steps, the
construction we have just sketched}.

The second section is a more mathematical one dedicated to a more special
presentation of the theoretical foundations of the ART. It does not contains
any physical result, but we thought it was necessary to ensure the global
consistency of ART. This second section is completed by an Appendix that
outlines the abstract theoretical conditions that a physical theory has in
any case to comply with and that stand behind those foundations.

The third section is dedicated to the first computations that result from
the MQT, giving first the aforesaid relation between Newton constant $G$ and
the fine structure constant $\alpha .$ As illustrative instances some
precise results concerning among other the ratios between well-known masses
of some particles are also presented here.

\bigskip

Far from being meant to give an achieved view of the Absolute Relativity
Theory, this paper has to be considered as a first proposal to the
scientific community to explore new ways that could lead contemporary
physics to some new results. It will certainly also lead to new questions
and difficulties that will have to be answered by a lot of complementary
work.

\subsection{The Absolute Relativity Theory in twelve steps}

Since the new approach we propose will lead to many constructions and
results that differ from those of contemporary theories, but are linked to
them, a short review may be useful for a first reading of this paper. This
is the purpose of this subsection. Precisions and justifications will be
given in the following sections.

\begin{enumerate}
\item  \textbf{The algebraic preuniverse }$\aleph$

\emph{The preuniverse $\aleph $ may be first chosen as the
category of complex finite dimensional $A$-modules with $A$
 any deformation of any complex quasi-Hopf algebra}. We will restrict
ourselves to Hopf algebras and to the quasi-Hopf algebras that they
generates by a gauge transformation. Those quasi-Hopf algebras may
be classified according to their primitive part $\mathbf{g}$. We will
furthermore restrict ourselves to Hopf-algebras with $\mathbf{g}$ a finite
dimensional semi-simple complex Lie algebra, beginning with simple ones. As
aforesaid, those Hopf algebras are generally not cocommutative : thus
instead of beginning as usual with the universal enveloping algebras, we will
consider modules ever Jimbo-Drinfeld Quantum Universal Enveloping algebras
(QUE), $U_{h}(\mathbf{g})$ as a starting point. We will restrict ourselves
to finite dimensional ones in order to apply the restricted duality to go from $
U_{h}(\mathbf{g})$ to its dual $F_{h}(\mathbf{G})$.

\item  \textbf{The intrinsic breach of symmetry in complex Lie algebras}

Since the ART is founded on a categorial framework that is defined up to an
equivalence, it is invariant under isomorphisms of Lie algebras. We may thus
choose any Cartan subalgebra $\mathbf{h}_{\max }$ of the maximal (in the
sense of inclusion) simple algebra $\mathbf{g}_{\max }$ (of rank $r_{\max }$
) we will work on (we will precise it below). We may then choose  
a Cartan subalgebra and an ordering of its 
roots to define the corresponding elements of any $\mathbf{g}_{i}$ of its
subalgebras (with $i\in I$, the finite set of Lie algebras we will work on, $
\mathbf{h}_{i}$ the corresponding Cartan subalgebra and $r_{i}$ its rank). 
\emph{Any irreducible object }$V$\emph{\ of }$\aleph $\emph{, as an }$
U_{h}(\mathbf{g}_{i})$\emph{-module or }$U_{h}(\mathbf{g}_{i})$\emph{
-representation functor }$\rho _{U_{h}(\mathbf{g}_{i}),V}$\emph{\ (or in
short }$\rho _{\mathbf{g}_{i},V}$\emph{), is characterized by a maximal
weight }$\lambda _{\mathbf{g}_{i},V}$\emph{\ and will be said to be at
\textbf{\ the position} }$V$\emph{\ (or }$\lambda _{\mathbf{g}_{i},V}$
\emph{)}.

So, like tangent spaces in General Relativity, our main objects will be
linear representation, but instead of using only the four dimensional
representations of the Lorentz group and seeing them as linked by a Lorentzian
connection, we will work on different QUE and their different finite
dimensional representations linked together by their algebraic intrinsic
relations that simultaneously define their relative characteristics and
(discrete) positions.

The set of all the $\lambda _{\mathbf{g}_{i},V}$ defines a set of lattices
all included in the dual of the Cartan subalgebra $(\mathbf{h}_{\max })^{*}$. We will call the lattice of weights $\Lambda _{i}$ associated with the algebra $
\mathbf{g}_{i}$ the $\mathbf{g}_{i}$\textbf{-type space-time} (in short $
\mathbf{g}_{i}$\textbf{-space-time}). \emph{It is not the usual
space-time, but it corresponds to the first step of our algebraic
construction of space-time from observables. As it is easy to check
in any Serre-Chevalley basis, although }$\mathbf{g}_{i}$\emph{\ is complex
algebra, each }$\Lambda _{i}$ \emph{is a copy of $(\mathbb{Z})^{r_{i}}$, and any }$\mathbf{g}_{i}$\textbf{-space-time }\emph{is thus an
essentially real object.}

\emph{So, although all the possible choices of a Cartan subalgebra} $
\mathbf{h}_{\max }$ \emph{are linked by inner automorphisms that are
complex ones, the set of finite dimensional modules defines in any
isomorphic copy of }$\mathbf{h}_{\max }^{*}$\emph{, the same real lattice.
Since a multiplicative coefficient }$e^{i\phi }$ \emph{does not preserve
this lattice, we have here an algebraically defined breach of symmetry in
the complex plan.}

This breach of symmetry is intrinsic in the sense that his existence does
not result from any postulated field such as, for instance, Higg's field,
but from the fact that \emph{finite dimensional representations of a
simple complex algebra are defined by a $\mathbb{Z}$-lattice}.

\item  \textbf{Working with both complex and real algebras}

We have thus to focus on real algebras, and we will proceed very carefully
by begining with the compact form canonically associated with each complex
algebra. Since any real form of any $\mathbf{g}_{i}$ is defined by a \emph{
conjugate linear} involutive automorphism, two real forms $\sigma $ and $
\tau $ are said to be compatible if and only if they commute, which means
that $\theta =\sigma \tau =\tau \sigma $ is a \emph{linear} involutive
automorphism of $\mathbf{g}_{i}$. Furthermore, since any real form is
compatible with the compact one $\sigma ^{c}$, the study of real forms comes  
down to the study the compact form and of its linear involutive
automorphisms.

Since any linear involutive automorphism of $\mathbf{g}_{\max }$ \emph{
that respects a subalgebra} $\mathbf{g}_{i}$, induces on $\mathbf{g}_{i}$ a
linear involution, any real form on $\mathbf{g}_{\max }$ induces a real form
on this subalgebra $\mathbf{g}_{i}$ ; reciprocally, if we identify a
specific real form $\tau _{i}$ associated with some linear involution $
\theta _{i}$ of an algebra $\mathbf{g}_{i}$ as associated with a family of
true physical objects, $\theta _{i}$ will have to be compatible (i.e.
commute) with the restriction to $\mathbf{g}_{i}$ of $\theta _{\max }$. The
same reasoning applies to other inclusions of algebras, and will determine a
lot of necessary theoretical conditions on compatible real forms of
algebras. Since the compatibility of real forms is a ``commutative
diagram''-type consistency condition for the theory, this way of reasoning
will be the key of the determination of the real algebras that define the
particles that can coexist in our universe.

\emph{Finally, the only real algebras we will be interested in, are the
compact ones, and only one family of other ones}.\emph{\ This leads to
distinguish in the Weyl chamber associated with the lattice of weights $\Lambda _{i}$
those that are compact and the other ones : the first
characterizes what we propose to call \textbf{folded dimensions}, the other
ones to\textbf{\ unfolded dimensions}. Any (different from }$\mathbf{sl}
_{2}\approx \mathbf{su}_{1,1}$\emph{) non compact real }$\mathbf{g}
_{i}^{\theta _{i}}$\emph{-space-time has unfolded dimensions that we will
identify with usual space-time, but also folded dimensions (that correspond
to its maximal compact subalgebra).}

Therefore, any real form $\mathbf{g}_{i}^{\theta _{i}}$ of the complex Lie
algebra $\mathbf{g}_{i}$ of complex dimension $n_{i}$ and rank $r_{i}$, has $
d_{i}$ unfolded dimensions, $r_{i}-d_{i}$ folded dimensions, $n_{i}-r_{i}$
dimensions that characterize the Lie algebra structure, but do not
correspond to any dimension of space-time : that is the reason why we
propose to call them \textbf{structural dimensions}.

\item  \textbf{A first algebraic classification of particles}

We are now ready to give the basic correspondence with usual particles

\begin{itemize}
\item  \textbf{the neutrino} corresponds to the algebra $\mathbf{su}_{2}$.
It has no unfolded dimension, and thus no intrinsic proper time (which
implies that they appear as going at the speed of light). Nevertheless, we
will see that neutrinos appear to us as having a mass that comes in fact
from the asymmetry of our proper time that we will describe below. This mass
could therefore be seen as a ``dynamical ghost'' mass.

\item  \textbf{the electron} corresponds to the algebra $\mathbf{so}_{1,3}$,
the realified algebra of $\mathbf{sl}_{2}(\mathbb{C})$. It has one folded and
one unfolded dimensions that supports its proper time. We will see how the
unique unfolded dimension is linked to the gauge invariance group $\mathbf{
SU(1)}$ that characterizes electromagnetic interaction. The splitting $
\mathbf{so}_{1,3}\approx \mathbf{su}_{2}\oplus \mathbf{su}_{1,1}$ has a very
special importance since it generates a one dimensional non rigidity that we
will characterize by the fine structure constant $\alpha $ seen as the ratio
between the inertial and the electromagnetic energies of the electron.

Another important fact is that the smallest representation of $\mathbf{so}
_{1,3}$ is four dimensional, which is the dimension of our usual space-time.
As an advantage, this allows us to have a complete view on its structure,
and thus to deduce its equations from experimental processes. As a drawback,
this has reinforced the idea that this space-time \emph{in some sense includes  
all the physical phenomena, which is a completely wrong idea as far as other
interactions (excepting gravitation) are concerned.}

Finally, the outer automorphism that corresponds to the symmetry between the
two (isolated) nodes of the $\mathbf{so}_{4}$ Dynkin diagram, induces at each
position the fugitive coexistence of a representation of $\mathbf{so}_{1,3}$
with the direct sum of two copies of $\mathbf{su}_{2}$ that corresponds to
two copies of the above algebra. Since the corresponding neutrinos, as
aforesaid, leave the electron almost at the speed of light, we will say that
the true physical phenomenon associated with this symmetry of the $\mathbf{so}
_{4}$ Dynkin diagram is an \textbf{anti-confinement phenomenon}.

\item  \textbf{the usual quarks} correspond to the algebra $\mathbf{so}_{3,5}
$ (or $\mathbf{so}_{5,3}$). It has one folded and three unfolded dimensions.
We will see how those three unfolded dimensions are linked to the gauge
invariance group $\mathbf{SU(3)}$ that characterizes the strong interaction. 
\emph{The well known triality, that comes from the group of automorphisms
of }$D_{4}$\emph{\ Dynkin's diagram which is }$S_{3}$\emph{, ensures the
coexistence of three copies of the different representations of the algebra at each position} : this very special
mathematical property explains the usual confinement of three quarks by
giving a precise mathematical sense to their usual three ``colors'', namely
the three isomorphic forms (standard, spin + and spin -) of the complex
algebra $\mathbf{so}_{8}$ that necessarily do appear together on any
position $V$ independently of any complex involution that defines the
considered real form of this algebra. The folding of those three in the
algebra $\mathbf{g}_{2}$ leads to protons and neutrons at the corresponding
position, as one could have guessed.
\end{itemize}

\emph{The important fact is that one can compute the characteristics of
all those particles from the well known characteristics of those algebras.
As aforesaid, we will give some instances of such computations in our last
section.}

\item  \textbf{The algebraic interpretation of the four known interactions}

From ARP, natural transformations between representation functors in the
theory should correspond to interactions, and reciprocally. Here is the
basic correspondence with usual interactions :

\begin{itemize}
\item  \emph{outer automorphisms of any complex algebra, that are
associates with the automorphisms of its Dynkin's diagram, correspond to
interactions} : the one of $D_{2}$ applied to $\mathbf{so}_{1,3}$
corresponds \textbf{to electromagnetism}, the triality of $\mathbf{so}_{3,5}$
to \textbf{strong interaction}.

\textbf{The weak interaction} will appear as corresponding to the outer
involutive automorphism that comes from the compact form the real form $
E(II)$ of the algebra $E_{6}$. We do not perceive it as such, but only
through ``ghost effects'', i.e. as representations of the particles we are
made of, which induces a splitting of the corresponding $E_{6}$-bosons (that
have to be massless) into different parts. Let us emphasize that such parts
can individually appear as massive, exactly in the same way a photon would
be perceived as a massive particle by one-dimensional observers with our
proper time. Bosons $W$ and $Z$ correspond to this situation. $E_{6}$ is also the ``largest'' exceptional Lie algebra that has outer automorphims, which is one reason why we can choose it as our $g_{max}$.  

\item  \emph{Weyl symmetries} applied to the space-time defined by an
observer correspond to changes of the Weyl chamber in the corresponding
weights diagram : we propose to call \textbf{changes of ``flavors''} the
corresponding relativistic effects, since, in the case of the algebra $
\mathbf{g}_{2}$ (that corresponds to our nucleons), the six
``flavors'' of leptons and quarks, associated with the corresponding
antiparticles, will appear as related to the twelve copies of the Weyl
chamber in this algebra. \emph{The important fact is, as above, that this
view of the ``flavor'' of particles gives another example of relativistic
effects that modify the characteristics of the observables according to
their position (in the general sense of ART) relative to the observer.}

\item  Since any element of any simple compact group $\mathbf{G}^{c}$ of
rank $r$ is conjugate to exactly one element of each of the $\left| W\right| 
$ sheets that cover its maximal torus, there is exactly one element $t$ in a
given copy of the Weyl chamber associated with any element of $\mathbf{G}^{c}
$. This allows us to see any element of $\mathbf{SU}(r)$ as acting on any
element of any specific sheet of the Weyl chamber, by keeping this element
if its image remains in the same sheet, or combining with an appropriate
Weyl symmetry, that appears as a change of ``flavor'', if it is not the
case. This allows us to encode everything in the maximal torus as explained
in subsection 1.3., and then to restrict ourselves to one of its sheets.
We propose to call \textbf{generalized Lorentz transformations} the pull
back of the so defined transformations through any specific involution that
defines any non compact real form.

There is also a possibility of working on any complex Cartan subalgebras of
the algebra $\mathbf{g}$ instead of the maximal torus of $\mathbf{G}^{c}$ by
substituting $\mathbf{sl}_{r}(\mathbb{C})$ to $\mathbf{SU}(r)$ and replacing
exchanges of sheets by exchanges of roots\emph{\footnote{{\small The way
ART is built legitimates the use of ``Weyl's unitary trick'' : we will use
it mainly to pull back any configuration in a Cartan subalgebra on the
maximal torus of the compact real form. Beginning with a postulated
space-time prohibits from applying the necessary involutions to go from the
compact real form to any other one. By contrast, in ART, this gives in
particular a way to make probabilist computations without having to deal
with infinites by some renormalization methods. Nevertheless, we will not
use those possibities in this first paper.}}}. A key fact is that those two
ways are equivalent since we will need both of them : the compact real form
in order to access to any other real form by an appropriate complex
involution, and the complex algebra in order to give sense to this
involution.

\item  \textbf{the gravitation} will appear as a direct consequence of the
passing of time as we will explain below.
\end{itemize}

\item  \textbf{Some algebraic conjectures on particles and interactions}

Our algebraic approach will also lead us to conjecture the existence of some
unknown particles and interactions :

\begin{itemize}
\item  The compact form of the $\mathbf{so}_{8}$ algebra should define
particles that are, as those associated with $\mathbf{so}_{3,5}$, linked by
the strong interaction induced by the mathematical triality. They should be
also folded into compact copies of representations of $\mathbf{g}_{2}$. We
thus propose to call them \textbf{dark quarks} that have to be folded into
what we suggest to call \textbf{dark neutrons\footnote{{\small We will
understand later on why there should not be white dark protons.}}}. Since
they correspond to a compact form, the space time associated with those
particles should not have any unfolded dimension, and they should be related
to usual quarks by an as above defined anticonfinement phenomenon. Dark
neutrons should also get a mass from the same relativistic effects as
neutrinos. Therefore, they could be part of the ``\textbf{dark matter}''
contemporary physics is looking for.

\item  The representations of the algebra $\mathbf{F}_{4}$ comes itself from
the folding of the outer automorphism that links the compact form of $
\mathbf{E}_{6}^{c}$ and the afore mentioned $\mathbf{E}(II)$ real form of $\mathbf{
E}_{6}$. Those three algebras should define particles and phenomena we do
not perceive as they truly are. As aforesaid, the weak interaction should be
one of them, but it is only a ghost effect.

To access to the true effect, it is necessary to represent the above outer
automorphism, and compute the corresponding Lagrangian as explained below.
This Lagrangian should define a new interaction that encompasses the weak
interaction. Furthermore, since both algebras $\mathbf{F}_{4}$ and $\mathbf{E
}(II)$ have real rank four, which explains the dimension of our usual
space-time, this new interaction should correspond to high valued
Lagrangians, which makes it a good candidate to be the main part of the
mysterious ``\textbf{dark energy}''.

On another hand, let us notice that, as made of smaller particles, we
perceive those particles not as they are but through \textbf{splitting
effects} that transform the rigid structure of any irreducible
representation of $\mathbf{F}_{4}$ into a direct sum of representations of
smaller algebras. Since ``branching formulas'' on those special algebras are
growing very fast with the maximal weights of their representations, this
splitting process will induce the appearance of huge numbers of splitted
particles. Since, furthermore, direct sums do not induce any rigidity, this
splitting process also explains the incredible freedom of their possible
arrangements.
\end{itemize}

\item  \textbf{The existence and meaning of Lagrangians}

The next step is to consider the impact of the point of view functor by
considering the algebra $F_{h}(\mathbf{G})$, the (restricted Hopf) dual of
the corresponding $U_{h}(\mathbf{g})$, that should define what we perceive
of all physical objects and phenomena.

Beginning as above with compact algebras, \emph{the first key point is
that, as we said before, all the representations of all the Hopf algebras }$
F_{h}(\mathbf{G}^{c})$\emph{\ are indexed by the maximal torus }$T_{
\mathbf{G}^{c}}$\emph{\ that is a copy of }$(T_{1})^{r_{\mathbf{G}}}$
(with $T_{1}$ the trigonometric circle).

Let us then notice that the group $\mathbf{SU}(r_{\mathbf{G}})$, that
corresponsal form of $\mathbf{SL}(r_{\mathbf{G}})$,
define the group of all the transformations of the maximal torus.

But only those that correspond to unfolded dimensions can generate
transformations that we can perceive. With above notations, we have thus
also to consider the smaller torus $(T_{1})^{d_{\mathbf{G}}}$ and the
smaller group $\mathbf{SU}(d_{\mathbf{G}})$. With the above definitions of
electromagnetism and strong interaction, we find here the usual gauge invariance
groups of the QFT, namely $\mathbf{SU}(1)$ and $\mathbf{SU}(3)$.

Now, any Lie algebra $\mathbf{sl}_{n}(\mathbb{C})$ with $n>2$, which compact
real form is $\mathbf{su}_{n}$, has a very specific property : by
identifying it through the Killing form with its dual, one sees that the
tensor product $\mathbf{\mathbf{su}_{n}\otimes su}_{n}$ contains, as for any
simple Lie algebra, a one dimensional representation of $\mathbf{su}_{n}$
(the image of the identity map, i.e. the Casimir operator of $\mathbf{su}_{n}
$), and the image of the adjoint representation of $\mathbf{su}_{n}$ (that
is antisymmetric). But, in the $\mathbf{su}_{n}$ (or $\mathbf{sl}_{n}(\mathbb{C}
)$) cases, and only in them, \emph{there is in }$\mathbf{\mathbf{su}
_{n}\otimes su}_{n}$\emph{\ a second irreducible representation of }$
\mathbf{su}_{n}$\emph{\ that is symmetric and thus defines a bilinear
symmetric form $\mathcal{L}^{\prime }$ on }$\mathbf{su}_{n}^{*}\otimes\mathbf{su}_{n}^{*}$ that may be identified with the help of the Killing form with an endomorphism of $\mathbf{su}_{n}$. Let us emphasize that Lagrangians do not act directly on the
Cartan subalgebra, but on the algebra of the endomorphisms of this algebra
seen as a vector space. Therefore, although the action of any element $a$ of 
$\mathbf{GL}_{n}$ on the Cartan algebra $\mathbf{h}$ of any Lie algebra $
\mathbf{g}$ respects $\mathbf{h}$ only globally - this corresponds to a
simple change of observer in usual physics -, Lagrangians as corresponding
through the Killing form of $\mathbf{su}_{n}$ to\emph{\ endomorphisms} of $
\mathbf{su}_{n}$ itself, will remain invariant : the action of $a$ for them
is the one of a simple change of base in $\mathbf{su}_{n}$.

Coming back to any simple algebra $\mathbf{g}$, an outer automorphism $\phi $
of $\mathbf{g}$ induces a change of the primitive roots, and thus of the
fundamental roots and Weyl chamber, that does not result from the action of
the Weyl group. It sends correlatively the connected component of the
identity of $Aut(\mathbf{g})$ to another one. This change in the roots
diagram induces an automorphism $\widetilde{\Phi }$ belonging to $Aut(
\mathbf{su}_{r})$ ; $\widetilde{\Phi }$ may be seen as above through the
Killing form of $\mathbf{su}_{r}$ as an element $\Phi $ of $\mathbf{su}_{r}
\mathbf{\otimes su}_{r}$. It splits therefore into irreducible
representations of $\mathbf{su}_{r}$ with a one dimensional projection on
the Casimir operator, which we will call \textbf{the free part of the
Lagrangian }$\mathcal{L}_{\Phi }$, and the above \emph{$\mathcal{L}_{\Phi
}^{\prime }$} that we will call \textbf{its interactive part}. The sum of
both parts will be the \textbf{Lagrangian of the interaction corresponding
to }$\Phi $.

Since it neutralizes the component that comes from the adjoint
representation of $\mathbf{su}_{r}$ that generates the above connected
component of the identity, \emph{the Lagrangian can be seen as
characterizing a scaling on }$\mathbf{su}_{r}$\emph{\ and a change of
connected component of the identity in }$Aut(\mathbf{g})$\emph{\
independently of any basis chosen for the Cartan subalgebra }$\mathbf{h}$
\emph{. This explains its additivity and its independence of the basis
chosen for representing }$\mathbf{g}$\emph{, and also of the choice of the Cartan subalgebra (since they are all conjugate) and finally of }$
\mathbf{g}$\emph{\ itself (except through }$r$\emph{\ and }$d$\emph{). and the Lagrangians it defines}

Therefore, the choice of coherent Cartan subalgebras we made in the beginning appears now only as a commodity for computations.

Furthermore, if an algebra is included in another one, the smallest defines
a degenerate bilinear form on the space-time associated with the first, but
the construction keeps making sense. \emph{This gives its main interest to
the Lagrangian.}

This unfortunately does not mean that it is easy to compute. Indeed, since
it is invariant under the action of the adjoint group and thus of the one of
the Weyl group, the computation of the Lagrangian, in the cases where it is
not reduced to its free part, has to include at least as many elements than
the cardinal of the Weyl group (which is $192$ for $\mathbf{so}_{8}$, $51840$
for $E_{6}$ and $1152$ for $F_{4}$ !). Although there are many symmetries to
be used, we will not try to make those computations here.

Since the algebra $\mathbf{so}_{1,3}$ has a two-dimensional Cartan
subalgebra, electromagnetism has only a free part easy to identify with the
usual one. The same is true for the gravitation that is, as we will see,
directly connected to the Casimir operator.

Another historically important, but singular aspect of electro-magnetism has
to be quoted here : it admits a true four dimensional representation, which
means that it can fully be represented in our space-time that is four
dimensional, but for other reasons. The successes of corresponding theories
have therefore sustained the idea that this space-time should contain
everything that concerns physics : this is the egocentric postulate.

The correct way to approach electromagnetism in our context is to see it as
associated with a two dimensional Cartan subalgebra (forgetting it is only
semi simple), which means a two dimensional space-time. The exchange of
perpendicular roots (that defines $\mathbf{so}_{1,3}$ from $\mathbf{so}_{4}$
) defines a rotation on the compact form that expresses the charge in a two
dimensional context : it is in fact the two dimensional version of Dirac
equation. Let $\lambda _{1,2}$ be the generator of this rotation. It easy to
check (for instance by using the complex representation) that the associated
(free) Lagrangian is given by $\lambda _{1,2}^{2}$. But, as we have said,
Lagrangians that comes from the Killing form are invariant under the action
of the Weyl group, that in our degenerated case corresponds to the group $
S_{4}$. The true Lagrangian is therefore $\sum\limits_{i,j=1,4_{i\neq
j}}\lambda _{i,j}^{2}$, that corresponds to the usual one with appropriate
choices of units.

Finally, let us notice that, essentially for simplicity of notations, we
have presented here all the elements in the non quantized situation. They
are easy to transpose to the quantized one.

\item  \textbf{The passing of time and the definition of physical observers}

\emph{The change to }$F_{h}(\mathbf{G})$\emph{\ induced by the point of
view functor has a second fundamental consequence}.

\emph{Beginning with the }$\mathbf{SU}_{2}$\emph{\ case, }$F_{h}(\mathbf{
SU}_{2})$\emph{\ admits, beside one dimensional representations, a family }
$\pi _{t}$\emph{\ (indexed by its maximal torus }$T_{1}$
\emph{) of infinite dimensional representations on the space of formal
polynomials with one variable that may be seen, through formal characters,
as the space of finite dimensional representations of }$\mathbf{SU}_{2}$
\emph{\ (with highest weight the degree of the polynomial)}. Let us
emphasize that this representation $\pi _{t}$ does not appear in the non
quantized case, and that, considered as a quantization of a Poisson-Lie
group, it is associated with the Weyl symmetry $w$ of $\mathbf{SL}_{2}(\mathbb{C
})$ that exchanges the compact and non compact roots (associated with $
\mathbf{SU}_{1,1}$) of its realified.

More precisely, those polynomials that correspond to $\mathbf{SL}_{2}(\mathbb{C}
)$ are polynomial in $(x+\frac{1}{x})$, the one that corresponds to the
standard two-dimensional representation $V$ ; $(x+\frac{1}{x})^{n}$
corresponds to $V^{\otimes n}$ and $x^{n}+\frac{1}{x^{n}}$ to the
representation with maximal weight $n$ ; $\pi _{t}$ has thus to be seen as
acting on the polynomials in the variable\footnote{{\small We use here the
standard representation of the algebra in order to get a one unit progression.
The same reasoning applied to the adjoint representation is more
physical since it does not exchange at each step bosonic and fermionic
representations. To use the square of X instead of X gives the transposition.
}} $X=(x+\frac{1}{x})$, an operation that is associated with acting on
tensorial powers $V^{\otimes n}$ of $V$.

It is easy to check\footnote{{\small We give a complete expression of those
representations in section II. References are in [V.C.-A.P.], pp. 435-439..}}
that, with the usual notation $\left( 
\begin{array}{ll}
a & b \\ 
c & d
\end{array}
\right) $ for the linear forms that define $F_{h}(\mathbf{G})$, $a$ reduces
the degree by $1$, while $d$ increases it also by $1$, and $b$ and $c$ leave
this degree unchanged. Furthermore, from the defining relations of $F_{h}(
\mathbf{G})$ (seen as morphisms of the $q$-plan with $q=e^{h/2}$), the
action of the commutator $[a,d]$ splits into a sum of multiples of $bc$, $cb$
and the identity, all three that keep the degree of the polynomial unchanged
: considered as acting on the space of representations, $[a,d]$ is
therefore, up to a scaling, equivalent to the identity.

Since we have seen that maximal weights define the positions of the objects, 
\emph{the action of the linear form }$d$\emph{\ may be seen as a move to
the right on the one-dimensional space-time induced by taking the tensorial
product by }$V$\emph{, which ``creates'' a new instant after the last one,
and ``moves'' to the right all the others. The action of the linear form }$a$
\emph{\ is reciprocal, inducing a move to the left and annihilating the
first instant. We get thus creation/annihilation operators that apply
directly to time.}

\emph{More precisely, in our case of }$\mathbf{su}_{2}\oplus \mathbf{su}
_{1,1}$\emph{\ that may be seen as corresponding to ``a pure time'' (it is
in fact the time of an electron), }$d$\emph{\ corresponds to a ``creation
operator'' of new (discrete) instants that induces an increasing move of all
the past instants on the time line : this is exactly a (discretely) mobile
point on the time line ; }$a$\emph{\ has exactly the opposite effect}.

The sum $qa+q^{-1}d$, that is the dual of the quantized identity $K_{q}$ of
the $q$-plan, induces therefore a double motion : the first one that
corresponds to the first coordinate in the $q$-plan, may be seen as the
motion of a fixed frame expressed in a moving one, the second as a motion of
a moving frame expressed in a fixed one. Exchanging left and right actions
exchanges those two situations as $K_{q}$ and $K_{q^{-1}}$, which is
perfectly consistent. $L_{q}=\frac{K_{q}-K_{q}^{-1}}{q-q^{-1}}$ that
quantized the usual generator of the Cartan subalgebra has an analog action.

\emph{In that sense, in this simple case, the passing of time may be seen
as a relativistic phenomenon induced by the point of view functor applied by
an observer belonging to the space generated by }$K_{q}$\emph{\ and }$
K_{q}^{-1}$\emph{\ on himself. Using Drinfeld-Jimbo QUE, the same
reasoning applies to the dual of the usual quantized Cartan generator.}

It is interesting to notice that making a tensor product by $V$ is
equivalent to applying the point of view functor from the point of view of $
V^{*}$ (seen as a representation of $\mathbf{sl}_{2}(\mathbb{C})$, thus
differing as a $\mathbf{sl}_{2}(\mathbb{C})$-module,from $V$ by the application
of the antipode $S_{h}$).

By identifying $V$ and $V^{*}$ through the symplectic invariant form, one
sees that the above operation of tensoring by $V$ is equivalent to applying
the usual algebraic flip that exchanges left and right operations, the
antipode and the contravariant $\hom $ functor : the present instant appears
so as a (reversed) point of view from the last past instant on always the
same $\mathbf{sl}_{2}$-type particle at the same position $V$. One therefore
can see the above equivalent of a moving frame as defined by successive
applications of the point of view functor on \emph{the same particle at
the position }$V$\emph{, that is thus seen as non moving from its own
point of view.}

\emph{Coming now to the general case }$F_{h}(\mathbf{G})$\emph{, the
existence of a surjective homomorphism (that comes by passing to the dual
from the principal embedding\footnote{{\small The choice of the principal
embedding is the only consistent one for a given a Lie algebra after
choosing a Cartan subalgebra, a set of roots and an ordering of them (that
defines the Weyl chamber). See for instance [A.L.O.-E.B.V.], pp. 193-203.}}
of }$U_{h}(\mathbf{sl}_{2}(\mathbb{C}))$\emph{\ into }$U_{h}(\mathbf{g})$)
\emph{\ from }$F_{h}(\mathbf{G})$\emph{\ to }$F_{h}(\mathbf{SL}_{2}(\mathbb{
C}))$\emph{\ canonically associated with any choice of a Weyl chamber
allows us to generalize the above construction after any choice of an
element of the maximal torus and an element of a Weyl chamber, which defines an
operator }$\pi _{t,w}$\emph{. }

\emph{Any such operator }$\pi _{t,w}$\emph{\ defines as above \textbf{
the proper time }associated with any }$\mathbf{g}$\emph{-type particle
that takes place in the space of successive powers }$V^{\otimes n}$\emph{\
of the standard representation }$V$\emph{. The dependence in this
definition on the choice of an element of the Weyl group (and thus a Weyl
chamber) justifies what we have said above on the action of this group as
changing the ``flavor'' of particles.}

Furthermore, the indexation of those representations by the maximal torus
corresponds to the above gauge invariance groups, the complete one $\mathbf{
SU}(r)$ and the observed one $\mathbf{SU}(d)$. The application of this last
group may be seen as transferring the time direction in any other one in the
Weyl chamber : this corresponds to what we have called above a generalized
Lorentz transformation\emph{. This gives the possibility to transpose the
above defined proper time direction in any other one in the Weyl chamber
extending correlatively the above definition from time to space-time.}

\emph{A \textbf{physical observer }may be now defined as the combination
of : }

\begin{itemize}
\item  \emph{a real type} $\mathbf{g}^{R}$, which means a complex simple
Lie algebra $\mathbf{g}$ characterized by its compact form $\mathbf{g}^{c}$\
and an involutive transformation

\item  \emph{a position defined by a maximal weight }$\lambda _{0}$\emph{
\ in }$\mathbf{h}^{*}$\emph{\ }the dual of the Cartan subalgebra of $
\mathbf{g}$,

\item  \emph{three elements that characterize the above principal
embedding of }$U_{h}(\mathbf{sl}_{2})$\emph{\ into }$U_{h}(\mathbf{g})$,
namely :

\begin{itemize}
\item  \emph{\ the choice of a copy of the Weyl chamber} by reference to a
first one arbitrary chosen : this is equivalent to the choice of an element
of the Weyl group (or to the choice of one of the$\left| W\right| $ copies
of the maximal torus by pulling back as above explained to the compact form)
;

\item  \emph{the choice of an element }$\nu $\emph{\ in the Weyl chamber}
(or $t$ in the above chosen sheet of the maximal torus by pulling back $v$
to the compact form) by reference to a first one arbitrary chosen ; this is
equivalent to \emph{the choice of an as above defined generalized Lorentz
transformation} that applies to an intially chosen roots diagram in order to
identify the direction defined by $\nu $ with the one of the principal
embedding of $\mathbf{sl}_{2}$ ;

\item  \emph{the choice of a free Lagrangian} $\mu ^{2}=(m_{0}+iq_{0})$. $
m_{0}$ defines the embedding of the unfolded part of the realified of $
\mathbf{sl}_{2}$, while $q_{0}$ the compact one. By convention, we take $
m_{0}>0$ for a physical observer, and we will say that $-\mu $ defines the
corresponding anti-observer. $m_{0}$ will be called \textbf{the mass} of the
physical observer and $q_{0}$ its \textbf{charge}.

\emph{Since }$m_{0}$\emph{\ corresponds to a scaling factor applied to
the dual of the algebra }$\mathbf{sl}_{2}$\emph{\ that contains the dual }$
\mathbf{h}^{*}$\emph{\ of its Cartan subalgebra, it defines a scaling of }$
\mathbf{h}^{*}$\emph{\ and thus the density relative to the dual of the
Killing form of integer values, which means successive finite dimensional
representations. In that sense, }$m_{0}$\emph{\ defines the speed at which
the time creation operator generates new instants. The higher it is, the
closer (in the sense of the dual Killing form) are the successively
generated instants, which is simply the generalization of the Einstein-Planck
relations.}

\emph{The existence of a mass (and a charge) associated with any physical
observer is no more a mysterious property of nature, nor something that
particles take out from a mysterious ambient field in order to ``slow'' them
: it is the purely logical consequence of two mathematical elementary facts,
namely the passing of time induced by the point of view functor as above
explained in }$F_{h}(\mathbf{SL}_{2})$\emph{, and the existence of a
uniquely defined (in a given root ordering) principal embedding of }$\mathbf{
sl}_{2}$\emph{\ in any simple Lie algebra }$\mathbf{g}$.
\end{itemize}
\end{itemize}

Let us emphasize that, by the main property that led us to their definition, 
\emph{Lagrangians are invariant under the action of generalized Lorentz
transformations that generalized usual changes of point of view of physical
observers. This means that all the physical observers that differ only
through such transformations will share the same Lagrangians.}

In usual context of physics, this explains the universality of Lagrangians.
In our context, this explains first why we perceive an ``existing universe''
(and therefore why physics has been so trapped by the egocentric
postulate)\emph{. Indeed, as being (up to ``flavors'' that appear like
generating different particles) associated with the same Weyl chambers of
the same algebras, our particles share the same Lagrangians despite their
relative moves, and therefore appear as sharing the same universe. This
universe is thus seen as independent of each of them and can be easily taken as the
general scene where the whole story is played.}

Another very important fact has to be emphasized here : \emph{there is an
elegant way to express }$\mu $ \emph{in a Lagrangian context : since }$\mu
^{2}$ \emph{is proportional to the Lagrangian and }$\mu $\emph{\ defines
the above ``speed'' }$\dot{z}$\emph{\ of the ``passing of time'' of this
observer, }$\frac{\partial \mathcal{L}}{\partial \dot{z}}$\emph{\ is
proportional to }$\mu $\emph{. The right--hand side of Euler-Lagrange
equations will simply be the expression of that fact in a context taking
also into account the principal embedding of }$\mathbf{sl}_{2}$\emph{\ in }
$\mathbf{g}$.

Let us notice finally that the creation/annihilation operators of a type of
particles seen as physical observers appear as the creation/annihilation
operators of their proper time. Since, in ART, space-time is defined from
particles, space is naturally expending while time is passing. \emph{
Therefore, the so-called second quantization corresponds exactly to the
quantization of the time, as the first one to the quantization of energy. It
induces a quantized structure of time and correlatively of space-time as we
will see now, beginning as usual with the one dimensional case}.

\item  \textbf{The structure of the time of a physical observer}

In order to explore the structure of space-time from the point of view of a
physical observer, one sees from the above construction of $F_{h}(\mathbf{g})$
representations, that we have in fact to describe the situation for the pure time
of a physical observer of type $\mathbf{su}_{1,1}\oplus \mathbf{su}_{1,1}$
(i.e. an electronic-type observer), then to extend it to a space-time by
inserting it in the Weyl chamber of a bigger algebra $\mathbf{g}$ and
finally to apply all the possible generalized Lorentz transformations.

Let us thus consider an electronic observer at the position $N$, which means
it is associated with the space $V^{\otimes N}$. This space splits into
irreducible representations ; \emph{the biggest one }$V_{N}$\emph{\
corresponds to the weight }$N$\emph{, with multiplicity }$1$\emph{\ and
may be seen as \textbf{the actual position of the observer}}. The other ones
have multiplicities given by the Newton coefficients in $(x+\frac{1}{x})^{N}$
. So defined and split, $V^{\otimes N}$ may be seen as the \textbf{universe
from the point of view of the considered observer at the instant }$N$.

In this construction, the universe at the instant $N_{1}<N$ was isomorphic
to $V^{\otimes N_{1}}$ that is \emph{not} the restriction to
representations of weights less than or equal to $N_{1}$ of $V^{\otimes N}$,
which corresponds to the past instants \emph{as seen from the present}.
This means that, as far as time itself is concerned, an observer cannot
access to the past without being influenced by $N-N_{1}$ successive
applications of the above mix of the flip and antipode operators combined
with the point of view functor.

Since the situation with $\mathbf{su}_{1,1}$ can, as above, easily be
transferred to the quantized case and to any other simple real Lie algebra $
\mathbf{g}$, this means that \emph{the past universe as seen from the
present is not the past universe as it was when it was the present}.

In other words, when we try to access the past through some experimental
processes (for instance by long distance astronomical observations), \emph{
we always access to representations of the past as seen from the present (}$
V^{\otimes N}$\emph{) and not as it was from the points of view of
successive instants (which leads to work on }$V_{1}\otimes ...\otimes V_{m}$ 
\emph{)}. Let us emphasize that by dualizing to Hilbert spaces of function
as in QFT, we get in the first case usual the Fock spaces of QFT while in the
second case, we have to represent the fact that past instants were not
exactly the same as present ones.

We will call the past as it is seen from the present as \textbf{the ghost
past}, and the past as it was \textbf{the true past}. Since ART defines
space-time from observables, the notion of space-time \emph{from the point
of view of a physical observer} is defined by some $V^{\otimes N}$ with $V$
the standard representation of the Lie algebra that characterizes the type
of the observer. In the above sense, space-time is thus a ``ghost
space-time'', and, since a physical observer ``knows'' only his own point of
view, it is doubtful that the notion of ``a true space-time'' from his point
of view could even make sense.

\emph{Nevertheless, both ``past'' are interesting from an experimental
point of view, since the study of the ghost past predicts what we perceive
from the past, and the study of the true past gives access to the true story
allowing us to compute or predict physical data that can also be compared
with measured values, which increases therefore the falsifiability of the
theory. }

In particular, we will see in our last section that it is theoretically
possible,\emph{\ by representing the true past}, to compute the
characteristics of all the particles and interactions from our actual point
of view. The same set of algebraic relations implies that all the
computations can also be done \emph{in the ghost past that defines what we
perceive today}, which means that \emph{all those characteristics
paradoxically appear as having been constant, despite their true variations
in the true past.}

We propose to call this law, inconceivable with the usual vision of an 
\emph{existing} space-time,\emph{\ t\textbf{he ghost constancy law of
constants} : it means the numerous contemporary observations that seem to
show that usual constants of physics are actually constant, are in fact
biased by the fact that the models used for those studies do not represent
the difference between the true past and the ghost past.}

In other words, as no observer can hope to determine directly his own motion
by looking to objects attached to himself, nobody can hope to determine
directly if the constants of physics are really constant without taking into
account and compensating the fact that all our past is in some sense moving
with us.

We are now quite far from the usual vision of a space-time as a smooth
preexisting Lorentzian manifold. In particular, the above process of
creation in a quantized way of ``new instants'' for all the physical
observers in all the space-time exclude to use any preexisting differential
structure. In fact, the only solid tool we have at our disposal in order to
represent space-time from the point of view of a physical observer is using
the creation operator itself applied to different (and different types of)
physical observers.

We will proceed by successive steps beginning with, as usual, the $
\mathbf{sl}_{2}$ case.

We know that the $t$-indexed representations $\pi _{t}(F_{h}(\mathbf{SL}_{2}(
\mathbb{C})))$ induce the move on representations of $U_{h}(\mathbf{sl}_{2}(
\mathbb{C}))$ that corresponds to the passing of time while the group of
rotations $\mathbf{SU(1)}$ generated by the standard Cartan element of $
\mathbf{su}_{2}$ acts on the torus $T$ that indexes the representations $\pi
_{t}$.We have to deal with this complex situation in a canonical way that
will allow us to generalize our results to other algebras. The idea is to
see the free linear space generated by the set of irreducible
representations as \textbf{a configurations space}\footnote{{\small See for
instance [C.K.] pp.451-455 for exact definitions, and then pp. 455-479 for
the KZ equation and Kohno-Drinfeld theorem. We refer also to [V.C.-A.P.] pp.
537-550 for the connection with Yang-Baxter equation.}} with $N$ dimensions
and to quotient by the symmetric group $S_{N}$ in order to have all the
representations at the same point $P$.

It is then possible to define an operation of parallel transport of the $2$
-tensors algebra $\mathbf{su}_{1}\otimes \mathbf{su}_{1}$, that contains the
above Lagrangian, according to Knizhnik-Zamolodchikov equation (KZ equation)
associated with some $2$-tensor $\mathbf{r}_{ij}$.

This operation is consistent, i.e. defines an appropriate flat connection
allowing us to define a monodromy group above $P$, if and only if $\mathbf{r}$
satisfies the classical Yang-Baxter (YB) equation with spectral parameters
(which corresponds to a ``commutative diagram''-type consistency condition
for the dodecagon diagram associated with an operator of this type).

We have now to determine the $2$-tensor $\mathbf{r}_{ij}$ that corresponds
to the passing of time of a physical observer. Since we have seen that the
above defined creation/annihilation operator that induces the passing of time is
the (quantized) identity which corresponds through the Killing form to the
Casimir operator, this $2$-tensor has to be build from the canonical
polarization of this operator, namely the canonical $2$-tensor $\mathbf{t}
_{ij}$ defined by :

\begin{center}
$\mathbf{t}_{V,W}=\frac{1}{2}(C_{V\otimes W}-$ $C_{V}\otimes
Id_{V}-Id_{V}\otimes C_{W})$.
\end{center}

This tensor is invariant by the gauge transformations, therefore its use is compatible with the categorial fondations of the ART.

In order to satisfy YB consistency condition, we have then to set :

\begin{center}
$r(z_{1}-z_{2})=\frac{\mathbf{t}}{z_{1}-z_{2}}$.
\end{center}

KZ equation may then be written for a function $f$ with values\emph{\ in
the true past} $V_{1}\otimes ...\otimes V_{N}$ :

\begin{center}
$\frac{\partial f}{\partial z_{j}}=\frac{h}{2i\pi }\sum\limits_{k=1_{k\neq
j}}^{N}\frac{\mathbf{t}_{jk}}{z_{j}-z_{k}}f$,
\end{center}

where the formal coefficient is introduced only in order to facilitate the
below formulation of Drinfeld-Kohno theorem.

Let us emphasize that, since we are in fact working on the realified of $%
\mathbf{sl}_{2}(\mathbb{C})$, the choice of the identity map that corresponds
to four applications of the rotation of roots induced by the outer
automorphism of $\mathbf{so}_{4}(\mathbb{C})$, and therefore of the Casimir
operator, is perfectly coherent with the (free) Lagrangian associated with
the electromagnetic interaction that we have above defined. This justifies
our choice of the definition of the electron and will ensure that in the
ART, like in the QFT, it will be possible to consider the charge as an
imaginary mass justifying the above given definitions of those quantities.

On another hand, we know that the Hopf algebra structure of $U_{h}(\mathbf{su%
}_{1})$ defines on each representation $V$ a canonical bilinear element of $%
\mathbf{su}_{1}^{*}\otimes \mathbf{su}_{1}^{*}$ that corresponds to the
action of the Casimir operator on this representation, which is a scaling by
the coefficient $C_{2}(V)$. It is given for the representation of $\mathbf{sl%
}_{2}(\mathbb{C})$ with maximal weight $N$ by the formula $C_{2}(N)=(N+1)^{2}-1$%
. This coefficient is also well known for other algebras in such a way that
the transposition to other algebras of our construction will be possible.

We can thus attach to any irreducible representation $V$ an element of $%
\mathbf{su}_{1}\otimes \mathbf{su}_{1}$ defined by $\frac{1}{C_{2}(V)}$.

Therefore, the change from representation $N$ to $N+1$ induces a change of
this canonical element of $\mathbf{su}_{1}\otimes \mathbf{su}_{1}$ according
to the ratio $\frac{C_{2}(N)}{C_{2}(N+1)}$.

\emph{The key point is the following : we may apply to this element the
closed parallel transport that defines the KZ monodromy getting a
coefficient given by }$e^{h\mathbf{t}/2}$ \emph{with }$\mathbf{t}$%
\emph{\ the above canonical }$2$\emph{-tensor and }$h$ \emph{the formal
coefficient in KZ equation.} \emph{We can therefore choose\footnote{%
{\small To be precise, we need to work with h a transcendental nmber in
order to respect the h-adic structure of Drinfeld-Jimbo algebras. We do not
adress this question here since we can replace the above choice by arbtrary
close transcendental values getting only a function arbitrary close to a
continuous function which is enough for the computations we intend to
propose.}} a specialization of the coefficient }$h$\emph{\ in
Drinfeld-Jimbo algebra such that this coefficient corresponds to the ratio }$%
\frac{C_{2}(N)}{C_{2}(N+1)}$.\emph{\ We get a Lagrangian (that in this
simple case is reduced to its free part) that by parallel transport in the
configuration space is a continuous function from one representation to the
next one.}

In an informal way, the ``jump'' of Lagrangian that is induced by the change
of intrinsic form at each change of representation can be ``absorbed'' by
the $R$-matrix effect of the flip that, as above explained, corresponds to
the successive applications of the point of view functor.

This means that when the passing of time induces a move from the position $N$
to $N+1,$ the quadratic form that has to be applied to define the
corresponding variation of the scalar coefficient in $U_{h}(\mathbf{sl}_{2})$
is$\frac{k^{2}}{C_{2}(N)}$ for some fixed positive coefficient $k$.
Therefore, the increase of this scalar coefficient that corresponds to the
passage from $N$ to $N+1$ is $u_{n}=\frac{1}{\sqrt{C_{2}(N)}}$.

Finally, the scalar coefficient after $N$ steps as seen from the point of
view of a physical observer, i.e. \emph{as pulled back from its dual after
application of the point of view functor}, is given by the corresponding
series $S_{N}=\sum\limits_{n=1}^{N}u_{n}$.

Now, in order to be able to represent a physical observer as stable on the
maximal torus, we have to normalize the bilinear forms we use in such a way
that, on the adjoint representation, where we know that the action on $%
\mathbf{h}^{*}$ goes on a $2$ by $2$ basis and the action on the torus has a 
$\pi $-periodicity$.$ \emph{We have thus to normalize the Killing form in
such a way that the image of }$\pi $\emph{\ is }$2$\emph{\ which gives a
coefficient }$\frac{2}{\pi }$\emph{\ by reference to a standard
orthonormal initial choice}. \emph{We will refer to this quadratic form as
t\textbf{he normalized quadratic form}. It is the one that should be used in
the computations of the MQT since it induces the stability of any of the }$t$%
\emph{-indexed representations of }$F_{h}(\mathbf{SU}_{2})$\emph{\ that
define the passing of the time from the point of view of the considered observer.%
}

Correlatively, we get a well defined sequence $h_{i}$ of values of $h$ that
we will call \textbf{the }$h$\textbf{-sequence}. \emph{Since we have seen
that Lagrangians characterize the outer automorphisms of the simple Lie
algebras, using an }$h$\emph{-sequence instead of a fixed value of }$h$%
\emph{\ is fundamental : otherwise, there would be small jumps in the
Lagrangians that do not correspond to any true physical phenomenon and that,
correlatively, would lead to the introduction of many corrective actions.}

Let us emphasize furthermore that the structure of the time of a physical
observer arises from this monodromic approach that corresponds to a
topological braided quasi-bialgebra usually called $A_{\mathbf{sl}_{2}%
\mathbf{,t}}$ \emph{that is not coassociative, but only
quasi-coassociative, and has the same product and coproduct than the non
deformed universal enveloping algebra.} It has thus the same algebraic
left-right symmetry than this latter algebra, which is a huge advantage, but
thinking the composition of duration as non associative is quite difficult,
and in any case not currently used by physics.

That is the reason why we have chosen to begin by working on Drinfeld-Jimbo
algebras that are coassociative. \emph{Furthermore, since by
Drinfeld-Kohno theorem\footnote{{\small See for instance [C.K.] page 460.}},
for any semi-simple Lie algebra }$\mathbf{g}$\emph{\ there exist a gauge
transformation }$F$\emph{\ and a }$\mathbb{C}[[h]]$-\emph{linear
isomorphism }$\alpha $\emph{\ that links }$U_{h}(\mathbf{g})$\emph{\ to }%
$(A_{\mathbf{g,t}})_{F}$\emph{, the representations of both
quasi-bialgebras define the same categorial structure, which ensures the
consistency of our whole construction that began precisely by those
categorial monoidal structures.}

\emph{There is nevertheless a price to pay to go from the one to the other
: we transform quasi-associativity in associativity, but we loose the
algebraic left-right invariance of }$A_{\mathbf{g,t}}$\emph{. We will
refer to this asymmetry as \textbf{the asymmetry of the time} of any
physical observer.}

\emph{From a physical point of view, it is the counterpart of having an
associative law for durations, and any precise calculation should be made
either on }$U_{h}(\mathbf{g})$\emph{\ taking into account this asymmetry
of time, or on }$A_{\mathbf{g,t}}$\emph{, but taking into account Drinfeld
associator that involves iterated integrals and Riemann }$\zeta $ \emph{%
function.}

\emph{Nevertheless, in this introducing paper, we will compute the
Anomalous Magnetic Moment of electron in the algebra }$U_{h}(\mathbf{sl}_{2})
$\emph{\ without compensating the asymmetry of its proper time, and use
the difference with experimental results to access directly to an evaluation
of this asymmetry.}

Let us come back finally to ``ghost'' space-time. It is easy to understand
where does the difference come from with true space-time we have just
described : the variations of the specializations of $h$ defined by the
above defined $h$-sequence that are necessary to have a continuous
Lagrangian are not taken into account since all the KZ story is told on the
same (initial) representation.

Unfortunately, contemporary physics seems to work exclusively on the
``ghost'' space-time from our point of view. This has tremendous
consequences since it forbids physics from accessing to its unification with
the understanding of the nature gravitation and to the Mass Quantification
Theory as we will see it below. Let us before come back to Euler-Lagrange
equations and their interpretation.

\item  \textbf{Recovering Euler-Lagrange equations}

We have already given in our subsection 1.3 above the main ideas in order to
understand the origin of Euler-Lagrange equations. So we only give here some
complementary elements :

\begin{itemize}
\item  We have seen that, if one thinks, in any given direction the operator 
$\frac{\partial }{\partial \dot{z}}$ applied to the Lagrangian gives its
free part that corresponds to a ratio of scaling in the corresponding
direction. We are thus always working on multiples of $1\otimes 1$. Since
the Casimir operator, and correlatively the canonical $2$-tensor makes
successively an action and its dual on anyone of the directions of the
considered Lie algebra, those variations are simply added if one uses an
orthogonal basis of the Lie algebra. But, since the canonical $2$-tensor $%
\mathbf{t}$ and KZ equations are defined intrinsically, and therefore
independently of the choice of any basis, the action of $\mathbf{t}$ makes
it possible to extend our way of reasoning in all the directions to be
considered.

\item  Remembering that any universal enveloping algebra of a Lie algebra $%
\mathbf{g}$ of simply connected Lie group $\mathbf{G}$ can be seen\footnote{%
{\small See for instance[S.H.], pp. 107-108.}} as the algebra $D(\mathbf{G})$
of operators on $\mathcal{C}^{\infty }(\mathbf{G})$ generated by the
identity and all the left invariant vector fields on $\mathbf{G}$. Let us
apply this to the group $\mathbf{SU}(d)$ (or $\mathbf{SU}(r)$ on $\mathbb{C}$).
The element of the universal enveloping algebra are now represented by
partial derivative operators that correspond to the usual expression of the
right-hand side of Euler-Lagrange equations.
\end{itemize}

\item  \textbf{The unification of physics, the gravitation and a conjecture
on dark energy }

As for the preceding step, we have already given in subsection 1.3 the main
aspects of this part, except as gravitation is concerned. Therefore, we give
only here some complements and precisions.

\begin{itemize}
\item  As far the unification of physics is concerned, let us first notice
that despite the name we propose, ART is clearly a quantized theory. Let us
emphasize that quantizations in ART comes from the contravariance of the
point of view functor since it obliges in the beginning to work on
bialgebras of which primitive part are well classified Lie algebras, and
since it gives rise to the fundamental above creation/annihilation operators
that sustain all the construction. In particular, we did not introduce any
physical hypothesis to get such a quantization.

On another hand, we do not present here any formal link with usual branches
of Quantum Theories. But, uncertainty principle for instance is easy to
deduce from the fact that a physical observer is only spectrally defined
despite the continuous view on itself obtained from KZ-parallel transport.
In the same way, the fact that exchanging algebraic right and left (i.e.
changing to the linear but not algebraic dual in non deformed algebras)
exchanges (roughly) time and energy directions should lead to derive easily
usual ``first'' quantization formulations that concern energy-momentum. We
have also sketched the way to get Dirac equation from the ART context.

There are also many connections with QFT we have quoted , but
we did not try to achieve formally the construction of a way to derive QFT
from ART.

\item  Far more interesting is the way we sketched in 1.3 in order to derive
Einstein equation from ART. The left hand of this equation is purely
mathematical, and directly results from the choice made by Einstein to use a
Lorentzian manifold in order to represent space-time. One could therefore
say that to derive from quantics a tensor that can be identified with usual
Lagrangian achieve the unification of physics since the right-hand side of the Einstein
equation is considered as given.

In fact we have had a little more. Since the variations from some given
point of our Lagrangians arise as a well defined KZ monodromy of the
symmetric part of an algebra of automorphisms of the Cartan subalgebra that
represents space-time in our context, and since the Ricci-curvature tensor
is its exact equivalent in a Lorentzian manifold context, both sides of
Einstein equation have now exactly the same mathematical status. Using, as
Einstein did, Euler-Lagrange equations for the representation of the
physical universe in the right side of Einstein equation appears now as an
useless detour : according to Einstein's hope, the true nature of his
equation is mathematical.

\item  As far as gravitation is concerned, let us first notice that the KZ
monodromy that we have described in order to define space-time from the
point of view of an observer has, from any representation $V$ to another
one, an action on $\mathbf{su}_{n}$ and thus a Lagrangian. As it comes from
an $\mathbf{sl}_{2}$-embedding effect, this Lagrangian has only a free part,
and his square root may be seen as defining a scaling of the Cartan
subalgebra itself. We have already analyzed this effect that corresponds,
from the definition of the KZ-parallel transport, to the change of the intrinsic
quadratic form associated with the change of the value of he Casimir
coefficient. This leads directly to the computation of the Newton constant
presented in section III.

\item  Coming back to Einstein equation, the fact that the gravitation
arises from a passing of time phenomenon that in any case defines a
Lagrangian, leads to conjecture that the right-hand side of Einstein equation would
not be modified by considering particles with an associated space-time of only two- or
three- dimension. But, the left-hand side would be modified by the
change of the unitary substraction that gives the projection on the pure
Ricci component in the space of Riemann curvature tensors. The law of
decrease of the gravitation would correlatively be slower (with a
logarithmic potential in dimension $3$, instead $\frac{1}{r}$ in dimension
four).

This phenomenon is probably not purely a theoretical one. Indeed, since, as
we will see in the MQT part, protons and neutrons because they are of type $g_{2}$ that as the normal form of $G_{2}$ has a rank of 2, cannot be described by small balls in space but by small bars in their own spacetime. Therefore, a concentration of them in one or two spatial dimensions if they are rotating seems to be conceivable
and could generate a gravitational field equivalent to the existence of some
``dark matter''. This remains nevertheless for us only a conjecture.

\item  Finally, let us remind the conjecture in 1.3 about ``dark energy''.
If one admits that, for the aforesaid reasons, weak interaction as we
perceive it, is only a ``ghost effect'' induced by some $E_{6}$-type
interaction, there is necessarily an (heavy) Lagrangian that corresponds to
it : a very good candidate for ``dark energy''. Furthermore, if such
particles and interactions exist their ``passing of time'' should generate a
sequence of increasing representations that split very fast in smaller
particles as the one we are made of. This leads to an alternative scenario
to the Big-Bang for the cosmological preradiative era... that we did not at
all explore.
\end{itemize}

\item  \textbf{The Mass Quantification Theory}

The way to compute the mass and the charge of particles is to compute the
free part of the associated Lagrangians, and, after taking the square root
that has a positive mass (for particles), to compare with corresponding
characteristics of the electron that we have chosen as a reference. 

In order to compute this Lagrangian, it is first necessary to position
rigorously the principal embedding of $\mathbf{sl}_{2}(\mathbb{C})$ that
defines the direction of the proper time of the considered particle : this
in fact a simple consequence of the way the creation/annihilation operator
that generates passing of time is positioned relatively to $F_{h}(\mathbf{G})
$ as above explained. Positioning the real and imaginary parts of the $%
\mathbf{sl}_{2}$ embedding has then to be made, either, if it is non
ambiguous, by identification of an unfolded direction, or by making the
embedding in the compact form and applying consistently the complex
involution that defines the considered real form.

The most difficult part is to compute how the passing of the time of the
considered particle has modified its free Lagrangian, since we do know
neither this evolution, neither the number $N$ of successive ``instants''
that correspond to this particle in its present proper time.

Fortunately, there is one exception : the electron that does not correspond
to a simple Lie algebra, but only to a semi-simple one. Therefore, the ratio
between its electromagnetic energy (that is orthogonal to its passing of
time, and thus does not change) and the one that corresponds to its mass,
that evolves as above explained, may be used to compute the present value of 
$N.$ This gives the apparent age of the universe (in its ghost past) and
permit, as aforesaid, the computation of Newton constant. As aforesaid we
will also use, in this first paper, the anomalous magnetic moment of the
electron in order to evaluate the impact of the asymmetry of time that
should be taken into account for other computations here. By contrast with $%
\alpha $ that characterizes our cosmological position, the anomalous moment
of electron should be computed from Drinfeld's associator in further
developments of the theory.

The idea for the other particles is then to use this value of $N$ to
determine the impact of their passing of time. To this aim, one has first,
by computing the above principal embedding and projecting on it the root
that corresponds to the adjoint representation of the considered algebra (or
its image through a Weyl transformation if the flavor is not the standard
one), to determine the relative speed of the passing of time of the
considered type of particle and compute its relative lag or advance by
reference to the speed of the adjoint representation of $\mathbf{sl}_{2}$,
and after a small conversion, relative to the one of the electron.

Once this lag or advance ratio is determined, one has to compute the
impact of this lag (for instance) relative to the electron. To this
aim, the idea is to consider the point of view of the electron looking to
the past, which means passing to the dual and correlatively representing the
annihilation operator from his point of view.

Using Yangians gives an easy way to make this computation since motions in
time can be represented by translations on the evaluations representations.
Now the derivation $D$ canonically associated with the Yangian $J$ (which is
obtained by applying successively the cocommutator and the Lie bracket on
the non deformed algebra) corresponds simply to $D=\frac{d}{du}$. Applying
this twice,since we move in the adjoint algebra, we get the equation that
has to be applied to any scaling to describe its evolution, namely :

\begin{center}
$\frac{d^{2}K}{(du)^{2}}=K$.
\end{center}

Since the initial value is $1,$ and since if one forgets the asymmetry of
the time, which implies a symmetric law, we get what we propose to refer as
the $\mathbf{cosh}$\textbf{\ law} : the coefficient to be applied in order
to compute the actual value of a particle is (in the lag case) equal to
the $cosh$ of the delay. Appropriate transformations have to be made in
cases of advance.

It is theoretically possible to improve those computations by adding a small
part that corresponds to the above explained asymmetry of the time, but we
do not present here any such improvement.

Although everything should be representable and thus computable, all those
difficulties lead us to present only some instances of such computations in
this first paper. This the purpose of our third section.
\end{enumerate}

It is time now to go further into details on more theoretical aspects of the
ART. After those twelve steps, we come back to the foundations.

\section{The mathematical foundations of the Absolute Relativity Theory}

Our objective in the first section was to outline the main
physical issues and perspectives of the ART. Therefore we only sketched its
mathematical foundations in the first steps of our overview in subsection
1.4.

Our purpose now is to give some complementary mathematical description of those
foundations. In order to facilitate an independent reading of this section,
we do not refer to elements given in the first section, but present here a
complete set of algebraic definitions and justifications.

\subsection{The algebraic approach}

Since we do not allow ourselves to postulate the existence of any
space-time, the first difficulty we have to overcome is to find a way to
begin with our theory.

According to Einstein's intuition quoted as epigraph, algebra soon appears
as the most natural way to undertake our quest.

\emph{We then begin to determine the algebraic conditions physics
has to respect to be a consistent and falsifiable science 
: with Kant's terminology in mind, we propose to call them 
\textbf{the transcendental conditions}. Those transcendental conditions are algebraic and define a very precise framework,
independent of any observer and of any observation, that also allows us to
represent the perception process itself as the action of a specific
contravariant functor.}

\emph{This general algebraic framework
has to be connected with ``the reality'' we perceive and that we suggest referring
to as the set of ``\textbf{true physical phenomena}''.}

\emph{With this aim in view, \textbf{we will state two and only two
physical principles}}

\begin{itemize}
\item  \emph{The first one, \textbf{the absolute relativity principle},
will appear as the ultimate extension of the above relativity principle.}

\item  \emph{The second one, \textbf{the absolute equivalence principle},
will appear as the ultimate extension of Einstein's equivalence principle,
and, in some way, as a reciprocal principle of the first one.}
\end{itemize}

\subsubsection{The preuniverse and the point of view functor}

In order to avoid beginning with any specific restriction to the theory we
are trying to build, it is natural to base the foundations of our
construction on the very general algebraic language of graphs. As
explained in Appendix I, Popper falsifiability conditions then immediately
lead us to focus on graphs that respect the axioms of\emph{\ categories :}
from now on, we will assume that the general framework of the theory we intend
to build is a category that we will call \textbf{the preuniverse} $\aleph $.
It will appear as a category of \textbf{representation functors}, namely
functors between two one-object categories, each one of them equipped with
morphisms that characterize its own internal structure. The morphisms in $%
\aleph $ will thus be morphisms of functors, i.e. \textbf{natural
transformations}.

It is then natural to define in each category\textbf{\ the point of view of
an object }$a$\textbf{\ on an object }$b$ as the set of arrows $Hom(b,a)$ ;
this defines \textbf{the point of view (contravariant) functor at }$a$, $%
Hom(.,a)$. Now, a functor $F$ is said to be \textbf{representable}\emph{\
and }\textbf{represented by }$a$ if there exists some object $a$ in the
target category of $F$ such that $F$ is equivalent to $Hom(.,a)$. Thus, the point
of view functor $Hom_{a}(.,a)$ will be the functor that sends $a$ to the set
of arrows that defines its algebraic morphisms as a one object category,
while $Hom_{\aleph }(.,\rho _{a})$ will be the functor that sends the
representation functor $\rho _{a}$ in $\aleph $ to the set of natural
transformations with the functor $\rho _{a}$ as target.

We sketch in Appendix I a way to show that some purely logical conditions on
the theory we try to build, such as being consistent in the sense of
categories and having ``enough'' representable functors, lead to specify a
little more the preuniverse $\aleph $ : \emph{it has first to be a
category of }$A$\emph{-modules (or ``representations of }$A$\emph{'')
for different algebras }$A$\emph{. Then, since each of the subcategory }$%
\aleph _{A}$\emph{\ associated with an specific algebra }$A$\emph{\ has
also to be a monoidal category for some coherent coproduct }$\Delta _{A}$%
\emph{\ and counit }$\varepsilon _{A}$\emph{,\ each }$(A,\Delta
_{A},\varepsilon )$\emph{\ has in fact to be a quasi-bialgebra\footnote{%
{\small See for instance [C.K.], pp. 368-371.}}. For technical reasons, we
will restrict ourselves to those quasi-bialgebras that are equipped with an
antipode operator, i.e. quasi-Hopf algebras.}

\emph{More precisely, in order to begin with the simplest of those
algebras and to use the restricted Hopf duality, we will focus on the
category of all the finite dimensional }$A_{h}(\mathbf{g})$\emph{-modules
for all the formal (}$h$\emph{-adic topological) deformations }$A_{h}(%
\mathbf{g})$\emph{\ of the universal enveloping algebras }$U(\mathbf{g)}$%
\emph{\ of all the finite dimensional semi-simple complex Lie algebras }$%
\mathbf{g}$. We will mainly use the well known Drinfeld-Jimbo Quantum
Universal Enveloping algebras (QUE) $U_{h}(\mathbf{g})$.

\paragraph{A first approach on particles and space-time}

Classically, those modules may be split into a direct sum of irreducible
submodules indexed by maximal weights that belong to the dual $\mathbf{h}%
^{*} $ of any chosen Cartan subalgebra $\mathbf{h}$ of $\mathbf{g}$. This
leads to the idea that \emph{those irreducible modules could correspond to
elementary particles, that arise, up to isomorphisms, as indexed by the
elements of maximal weights included in }$\mathbf{h}^{*}$.

\emph{This would mean that space-time does not preexist, but is defined as
a set of indexes by each type of particles, and in particular that its
dimension is defined by the rank }$r$\emph{\ of the Cartan subalgebra that
corresponds to their type.}

Now, since the lattice of maximal weights is linearly isomorphic to $(\mathbb{Z}%
)^{r}$, \emph{any choice of a Cartan subalgebra induces a breach of
symmetry in the complex algebra }$\mathbf{h}^{*}$\emph{\ that defines a
specific embedding of }$\mathbb{R}$\emph{\ in }$\mathbb{C}$. Since everything in
a theory is done up to an isomorphism, we will assume in all what follows
that such a choice is made for the biggest algebra we will have to consider,
namely, the algebra $E_{6}$, and, except when otherwise specified, we will
always consider that the Cartan subalgebra associated with any subalgebra of 
$E_{6}$ is chosen as the restriction of the one of $E_{6}$. We will apply
the same rule for the linear form that defines the ordering of the weights.

Since the above definition of space-time induces a specific embedding of $%
\mathbb{R}$\ in $\mathbb{C}$, we will have to work on real forms of Lie algebras.
Each of those algebras has a real rank $d$ that defines what we suggest
calling its\textbf{\ unfolded dimension}, and a complex rank $r$, the
difference $r-d$ corresponding therefore to\textbf{\ folded dimensions}.

Therefore making the definition of space-time dependant of each type of  particles instead of postulating them in
an \emph{a priori} given space-time independant of them should make us able to understand why
our space-time does appear as four-dimensional.

Let us define \textbf{an elementary particle of type }$\mathbf{g}$\textbf{\
at the position }$V$ as the irreducible representation functor of $U_{h}(%
\mathbf{g})$ on the one object category $V$, itself generally characterized
by a maximal weight $\lambda _{V}$.

\emph{For instance, the electron will be defined as the realified
of the algebra }$\mathbf{sl}_{2}(\mathbb{C})$\emph{\ that has a two
dimensional real Cartan subalgebra. Then the space-time of the electron will
appear as only one-dimensional since the compact root will correspond to an
unfolded dimension. We will see that our usual time is in fact the time of
the electron, and that it arises algebraically equipped with a both spectral
and continuous structure, far away from the copy of }$\mathbb{R}$\emph{\ that
usually represents the time in contemporary physics}. Any bigger algebra $%
\mathbf{g}$ with real rank $d$ should in the same way generate a $d$%
-dimensional space-time.

Concerning real forms of Lie algebras, let us remind\footnote{{\small See
for instance [A.L.O.-E.B.V.],, chapter 4, sections 1 to 4, pp.127-162.}}
that each of them is defined as an invariant\emph{\ conjugate linear}
involution of the corresponding complex algebra. The compact real form $%
\mathbf{g}^{c}$ is unique up to isomorphisms and may be used to define all
the other real forms of $\mathbf{g}$ : namely each one can be obtained by
applying an invariant\emph{\ linear complex} involutive isomorphism $%
\theta $ (that can clearly not be a real one) to the compact form.
Furthermore, each real form is characterized by the way $\theta $ acts on
its primitive roots, which can be summarized in\emph{\ a Satake diagram.}

Finally, let us notice that Algebraically replacing the usual left action by
the right one defines a second way to go from a set equipped with an
algebraic structure to a one-object category : we will say that those two
ways are \textbf{left-right conjugates}, and refer to the algebraic
operation that exchanges algebraic left and right actions as the \textbf{%
left-right conjugation }that will play a very special role in the absolute
relativity theory. Except when otherwise specified, the left convention will
always be used.

This conjugation that corresponds to the change to \textbf{opposite }%
algebras (symbol $op$) or coalgebras (symbol $cop)$ generally corresponds to
the switch from one representation to\emph{\ the linear dual representation%
}.

\paragraph{The dual category $\aleph ^{*}$}

\emph{The left-right conjugation that leads to the use of the linear
duality has to be clearly distinguished from the algebraic operation of
``reversing the arrows'' that is involved by the contravariance of the point
of view functor : in the latter case, everything is replaced by its dual,
which means, in the case of the above bialgebras, that coproducts are turned
into products and (with the restricted dual notion) products into coproducts}%
.

As usual, we will note the restricted dual of the algebra $U_{h}(\mathbf{g}%
)$ by $F_{h}(\mathbf{G})$. We will work on the dual algebra to
describe and explore the properties of the coalgebras and comodule
structures of the bialgebras and bimodules structures involved in our
construction. $F_{h}(\mathbf{G})$ is classically a quantization of the space 
$F(\mathbf{G})$ of complex valued functions on the simply connected Lie
group $\mathbf{G}$, and Tanaka-Krein duality allows us to see this group as the
dual of $F(\mathbf{G})$ seeing $U_{h}(\mathbf{g})$ as the quantum
group associated with $\mathbf{G}$.

The contravariance of the point of view functor will make 
$F_{h}(\mathbf{G})$ play a very central part. \emph{Their
representations are indexed by elements of the Cartan subalgebra
the duali of which is seen above as a support of the space-time}.

\emph{Furthermore, we some of those representations are defined on the
space of one-dimension formal polynomials, a space which is isomorphic to
the space of the representations of the algebra }$\mathbf{su}_{1,1}$\emph{%
, the non compact component of the realified of }$\mathbf{sl}_{2}(\mathbb{C})$,%
\emph{\ the algebra we already guessed characterizing the electron. Now
this algebra has a one-dimension non compact Cartan subalgebra, a natural
support for a one dimensional space-time, i.e. a time. The change to the
dual induced by the contravariance of the point of view functor leads to
associate the electron with this polynomial representation that, as we have
seen in section I, appears to be the origin of the passing of time. Our time
therefore corresponds to the time of electrons (and thus of
electromagnetism).}

Finally, having renounced to postulate any specific structure for the
space-time leads us to represent it with a far more sophisticated structure
than anyone we could have \emph{a priori }postulated.

\subsubsection{The ART two physical principles}

At this point, \emph{the preuniverse is nevertheless a pure mathematical
construction that needs to be linked to physics through the representation
of the phenomena we perceive.}

The refutation of the egocentric postulate gives here its first dividend :
since the structure of the preuniverse is not breached by any specific point
of view, one can guess that \emph{up to the application of the point of
view functor, any natural transformation in the preuniverse should
correspond to a physical phenomenon, and, reciprocally, that any physical
phenomenon should correspond to a natural transformation in this preuniverse}%
.

The above point of view functor would then associate any $\mathbf{g}$ type
particle at the position $V$ with all the true physical phenomena that have
an action on it, which corresponds to a natural representation of the usual
concept of ``the perceived physical phenomena''. Finally, \textbf{physical
laws} should be defined as families of natural transformations defined in
some universal way.

Those two reciprocal ideas are expressed respectively by the absolute
relativity principle and by the absolute equivalence principle.

\paragraph{The absolute relativity principle}

Let us notice first that the contemporary way to express the old idea of
``formal invariance of physical laws under a change of observers'' is to say
that such a change has to be defined by a natural transformation (the
adjoint action of Galileo or Poincar\'{e} groups for instance).

Now, the (classical) relativity principle defines a family of natural
transformations that link together Galilean observers, which effectively
induces a true physical phenomenon : the conservation of the
energy-momentum. In the same way, Einstein's relativity principle induces
the equivalence of Lorentzian observers, which implies also a true physical
phenomenon : this is Einstein's famous equivalence between energy and mass.

From those two examples one sees that the first of the above guesses
generalizes the idea that \emph{the compatibility of a natural
transformation (like a change of observer) with the theory is significant of
a true physical phenomenon} : we therefore refer it as \textbf{the absolute
relativity principle (ARP) }\emph{: any natural transformation in the
theory represents a true physical phenomenon.}

\paragraph{The absolute equivalence principle}

With his equivalence principle, Einstein went in the reciprocal direction by 
\emph{inducing}, from the numerical identity of the gravitational mass and
the inertial one, that the gravitation,\emph{\ as a true physical
phenomenon}, can be cancelled by extending the category of observers to
accelerated ones. This means that gravitation may also be represented by
natural transformations in this extended category.

Mathematically, this assertion has a direct consequence on the structure of
the space-time that has to be a differential manifold equipped with a
connection (necessarily Lorentzian in order to respect locally special
relativity theory).

This equivalence principle that ``cancels'' a physical
phenomenon by applying a natural transformation that embeds the category of
Lorentzian observers in a bigger one (or equivalently the category of affine
spaces in the bigger one of affine manifolds), we will be refered to 
\textbf{\ the absolute equivalence
principle (AEP)} : \emph{any true physical phenomenon may be represented
by a natural transformation of the theory.}

\paragraph{The structure of the absolute relativity theory}

\emph{Basically, taken together, the ARP and the AEP splits the
representation of the true physical phenomena into two parts :}

\begin{enumerate}
\item  \emph{the description and classification of the representations
functors and natural transformations that are the objects and morphisms of
the preuniverse }$\aleph $,

\item  \emph{the analysis of the impact of the point of view functor on
those objects and morphisms, which leads to work simultaneously on }$\aleph $%
\emph{\ and }$\aleph ^{*}$\emph{.}
\end{enumerate}

\emph{By classifying all those representation functors and natural
transformations between them, and combining this with the application of the point of
view functor, we should thus get a natural classification of the physical
particles and physical phenomena, which corresponds to the unification of
all particles and interactions\textbf{.} As we have seen in section I, this
physical unification comes with the theoretical one of the two main branches
of contemporary physics.}

\emph{As mentioned above, this also leads to\ the first Mass
Quantification Theory (MQT) that should explain the existence, the nomenclature
and the characteristics of the particles already identified by physics, and
also predict the characteristics of some new ones.}

Cutting the Gordian knot of space--time opens thus a new road to new
landscapes we are now ready to explore.

So, let us now begin with the first of the two steps above.

\subsection{A first classification of true physical phenomena}

\subsubsection{The fundamental structure of the preuniverse $\aleph $}

The preuniverse $\aleph $ may be seen as ``encapsulating'' three levels of
``arrows'' and universal constructions.

The first level encapsulates the basic transcendental conditions presented
in Appendix I and leads to consider $R$-modules for some ring $R$.

The second level encapsulated in the different representation functors $\rho
_{A,V}$, defines monoidal structures (see also Appendix I) on the category
of those modules, and induces the possibility to classify them according to
the different (complex and real) Lie algebras $\mathbf{g}$ that correspond
to the primitive part of different Hopf algebras $U_{h}(\mathbf{g)}$. This
level leads to the representation of particles and correlatively to the MQT.
In that sense, the existence of a well defined nomenclature of particles is
the consequence of the fact that, as explained in Appendix I, we want the
theory to have ``enough'' objects to ensure that the composed functors
defined by successive applications of the point of view functor, is
representable.

The third level corresponds to morphisms between those representation
functors, i.e. to natural transformations that will define, as aforesaid,
true natural phenomena that appear to any particle at any position through
the point of view functor.

\emph{Any natural transformation between two representation functors may
thus be seen as reflecting a (generally partial) formal coincidence between
two such structures of successive arrows and universal constructions that
``exist'' independently of any other point of view functor.}

\emph{As true physical phenomena are usually characterized by the fact
that they do not depend on any point of view, this explains why we have
chosen to associate them with natural transformations accordingly to the
above relativity and equivalence principles.}

\subsubsection{Objects and morphisms in the preuniverse $\aleph $}

To pursue our road, we have now to go into more details with the description
of the preuniverse.

\paragraph{Splitting the category $\aleph $ in subcategories $\aleph _{
\mathbf{g}}$}

First, if we specify different simple complex Lie algebras $\mathbf{g}$, 
\emph{we get different and distinct subcategories of functors we will
denote }$\aleph _{\mathbf{g}}$. Furthermore, since we have to work on the
real forms of those algebras, the same classification in subcategories
applies to those real algebras.

In particular, \emph{the classification of elementary particles by their type arises
this way : the leptons correspond (up to relativistic effects that define
their flavor as we will see later) to the real forms of }$\mathbf{sl}_{2}(
\mathbb{C})$\emph{, the neutrinos to the compact form and the electrons to
the split one, while usual quarks will appear as representations of a
specific real form of }$\mathbf{so}_{8}(\mathbb{C})$\emph{.}

\emph{Clearly, this way of seeing the particles should enable us to
predict the existence of so far undetected particles. For instance, since
the compact form of a real algebra is compatible with any other real form,
the compact form of the algebra }$\mathbf{so}_{8}$\emph{\ should also
correspond to physical particles that should go by three as the quarks and
that we propose to call \textbf{dark quarks.} We propose correlatively to
call\textbf{\ dark neutrons }the particles associated with their folding in }
$\mathbf{g}_{2}$.

\emph{Those particles are indeed good candidates to be elements of \textbf{
the unknown dark matter} contemporary physics is looking for. }

\emph{We will also be able to predict the existence of heavier particles
(associated with the algebra }$\mathbf{E}_{6}$\emph{\ and its folding in }$
\mathbf{F}_{4}$\emph{) that we see only through the explained
splitting and ghost effects induced by the fact that we see them from
the point of view of the lighter particles we are made of. One could
conjecture that those heavy particles originate \textbf{the dark energy}} 
\emph{according to a process we will try to describe at the end of our
second section.}

\subsubsection{The natural transformations between $\mathbf{g}$-type
representation functors : different physical effect}

\emph{Let us now classify the morphisms in any }$\aleph _{\mathbf{g}}$
\emph{\ of the above subcategories of functors }(with $A=U_{h}(\mathbf{g}) 
$ in order to simplify notations) :

Since each element of $\aleph _{\mathbf{g}}$ is a functor $\rho _{A,V}$
between the one-object categories $A$ and $V$, a natural transformation may
be :

\begin{itemize}
\item  either \textbf{type I}, induced by \textbf{an intertwining operator}.
Each one is defined by a \textbf{morphism of representation} $\phi
:V\rightarrow W$ that connects the two representation functors $\rho _{A,V}$
and $\rho _{A,W}$ by the relation : $\phi \circ \rho _{A,V}=\rho _{A,W}$. As
a natural transformation, each one is a morphisms of $\aleph _{\mathbf{g}}$.

Furthermore, if $V$ and $W$ are irreducible, Schur's lemma implies $V\approx
W$ and that $\phi $ is a scaling. An important consequence of this fact is
that, since any two representations of $\mathbf{su}_{1,1}$ at different
positions are never isomorphic, two distinct representations (or \textbf{
instants}) of $\mathbf{su}_{1,1}$ cannot be related by any natural
transformation, which explains why, as it is well known, from any specific
point of view, all the past instants do not exist from the point of view of
the present one.

Thus, \emph{if we restrict to irreducible representations, type I physical
phenomena can only be automorphisms of any given representation.}

From the universality of $U(\mathbf{g})$ in the category of the
representations of $\mathbf{g}$, one may deduce that the elements of its
center are the only ones (excepting trivial scaling) that induce a natural
transformation on each representation functor. We will be mainly concerned
by the Casimir operator $C_{h}$ that generates this center and defines the
coefficients of the intrinsic duality on $U_{h}(\mathbf{g})$-modules, and
that will appear as strongly connected to the passing of time and the
gravitation.

\item  or \textbf{type II }to \textbf{type IV}, induced by automorphisms of
the Hopf algebra $A=U_{h}(\mathbf{g})$. Since they have to preserve the
primitive part of this algebra, those automorphisms are associated with
automorphisms of the Lie algebra $\mathbf{g}$. They can come from outer
automorphisms of $\mathbf{g}$ (\textbf{type II physical phenomena}), or by
inner automorphisms that, for a fixed choice of the Cartan subalgebra, are
the product of a an element of the Weyl group of $\mathbf{g}$ (\textbf{type
III physical phenomena}), and a transformation that preserves the Weyl
chamber (\textbf{type IV} physical phenomena or ``\textbf{generalized
Lorentz transformation''}). This classification first applies to complex
algebras and has then to be specified by considering real forms of those
algebras.

Let $\mathbf{G=Aut(g)}$ the group of the automorphisms of $\mathbf{g}$ : it
is isomorphic to the adjoint Lie group associated with $\mathbf{g}$, and may
be obtained as a quotient of its universal covering that is simply connected. Let $\mathbf{G}_{0}
\mathbf{=Int(g)=\exp (g)}$ the group of inner automorphisms of $\mathbf{g}$
that is also the connected component of the identity in $\mathbf{G}$.
Classically, the quotient group $\mathbf{G}/\mathbf{G}_{0}$ defines the
group of the outer automorphisms of $\mathbf{g}$. It is isomorphic to the
group of the isomorphisms of the Dynkin diagram of $\mathbf{g}$, and each of
its elements corresponds to a connected component of $\mathbf{Aut(g)}$. By
contrast with the Casimir operator that acts differently on each
representation functor, automorphisms of $\mathbf{g}$ act in the same way on
all its representations and thus \emph{can only define ``relativistic
effects'', namely effects that link a representation to another one : they
thus define true physical phenomena that connect the corresponding particles.
}

Let us examine the main characteristics of each of those three types of
natural transformations :
\end{itemize}

\begin{enumerate}
\item  The existence of outer automorphisms (type II) implies that \emph{
with each }$\mathbf{g}$\emph{-module }$V$\emph{\ are necessarily
associated as many non connected copies of itself as there are such
automorphisms}. Furthermore, the mathematical operation of ``folding'' of
those copies that defines a specific subalgebra $\mathbf{g}^{\prime }$ of $
\mathbf{g}$, induces on the same space $V$ a $\mathbf{g}^{\prime }$-module
structure.

The first example of type II physical phenomenon comes from the triality\footnote{
{\small We refer here to the true mathematical triality that corresponds to
outer automorphisms of the $so_{8}$ algebra, and not to the one usually quoted in
physics and associated to $su_{3}$, that refers only to order 3 elements of the
Weyl group of this algebra. See the next subsection for an explanation of
the role of the algebra $su_{3}$ instead of $so_{8}$ in the usual interpretation of the
strong interaction.}} defined on the algebra $\mathbf{so}_{8}$, that links together through
natural transformations on any position $V$, the three fundamental
isomorphic representations of $\mathbf{so}_{8}$ (standard, spin$^{+}$ and
spin$^{-}$). \emph{Such representations appear to be good candidates
corresponding to the three ``colored'' quarks}, while the natural
transformations that connect those functors should correspond to the strong
interaction that ensures the confinement of those quarks into \emph{
nucleons that should therefore correspond to representations of the algebra}
$\mathbf{g}_{2}$.

\emph{There should be another couple of type II physical phenomena,
that are still unexplored.\ Namely, }$S_{2}$\emph{-type automorphisms of
Dynkin's diagrams induce another type of confinement, we propose to refer to
as \textbf{the anticonfinement} since it links for instance electrons and
neutrinos but in a repulsive way.}

\emph{Those automorphisms also imply the existence of new particles such
as the afore mentioned\ dark quarks that come from the compact form of }$D_{4}$
\emph{-representations folded in the above mentioned dark neutrons.}

\emph{Furthermore, we will see that there is also a type II physical
phenomenon that concerns }$E_{6}$\emph{-particles with an }$F_{4}$\emph{
\ folding we do not perceive directly.}\textbf{\ }

Finally, since we will also be concerned with real representations, let us
notice that, since it is an involutive automorphism, any $S_{2}$-type
automorphism of a Dynkin's diagram induces a modification of the real form
of the algebra well defined by the corresponding compact form.
This is the reason why, for instance, we mentioned above the anticonfinement
of electrons and neutrinos and not of same type particles. We will refer to
those changes of types of particles as \textbf{Type IIr physical phenomena}.
In practice, those phenomena correspond to the arrows in the Satake diagram
that defines each real form.

\item  \emph{Type III physical phenomena, like type II, correspond to
automorphisms of the Lie algebra }$\mathbf{g}$\emph{, but instead of outer
automorphisms, we consider now elements of the inner automorphisms group }$
\mathbf{G}_{0}$\emph{. By contrast with type II, their action preserves
each connected component of the adjoint group }$\mathbf{Aut(g)}$\emph{.}

It is known that the group of automorphisms of any (complex) Lie algebra $
\mathbf{g}$ is generated by the three dimensional algebras $\mathbf{g}
_{\alpha _{i}}$ in the adjoint representation where $\alpha _{i}$ runs on all the roots of a root system
of $\mathbf{g}$, and that the Weyl group $W$ acts simply and transitively on
the system of simple roots to get all the root system of $\mathbf{g}$.

\emph{Our theory of the passing of time
implies that the definition of an observer is biunivocally associated with
the choice of a system of simple roots that defines an }$\mathbf{sl}_{2}$
\emph{-principal embedding}. Thus, \emph{from the point of view of an
observer}, another observer associated with the same roots system appears as
the image of the first one through an element of $W$. This group therefore
defines a new set of natural transformations that, although non kinematical,
depend on the choice of an observer, and thus correspond to purely
relativistic effects.

We will see that \emph{the so called ``flavour'' of leptons and quarks
correspond to type III physical phenomena associated with the group (}$G_{2}$
\emph{) that defines the protons and the neutrons which we are made of.}

As for type III physical phenomena, the change from complex numbers to real
ones oblige to be careful since the composition of the conjugate linear form
that defines a real Lie algebra with an involutive transformation changes
the real form itself. Furthermore, the central Weyl symmetry always
corresponds to the change to the dual representation, a fact with many
consequences in the computation of the actual characteristics of the
concerned particles in the framework of the MQT presented in the section
III. When necessary, we will thus specify type III physical phenomena by
adding the suffix \textbf{r} (for real).

\item  Finally, since the $W$-orbit of any element in the dual of the Cartan
subalgebra intersects a closed Weyl chamber exactly once, we may always, by
an appropriate type III transformation, send the $\mathbf{sl}_{2}$-principal
embedding that defines, as we will see below, the proper time of an observer
into the Weyl chamber associated with another one. This operation eliminates
the ``flavor'' of the second particle relatively to the first.

We may then apply a change of basis that sends the set of primitive roots
that defines the first $\mathbf{sl}_{2}$-principal embedding onto the
corresponding one defined by the second principal embedding. Since the two
systems are normalized by the Killing form of the algebra $\mathbf{g}$, this
transformation has to be an orthogonal one.

We will say that the corresponding physical phenomenon is \textbf{a purely
kinematical effect}, or \textbf{a relative speed effect}, or else, since the
first instance of such a phenomenon is clearly the Lorentz group, \textbf{a
generalized Lorentzian effect}. Those natural transformations that arise in
this way will be said to be \textbf{type IV natural transformations }or
\textbf{\ generalized Lorentz transformations.}

\emph{We see here how restrictive the usual notions of relativity were :
the kinematical effects that were the only ones concerned by those usual
notions are now reduced to\textbf{\ residual effects }that play their part
only after all the other ones.}
\end{enumerate}

\paragraph{The natural transformations induced by the inclusion functor
between algebras}

Having listed the nomenclature of natural transformations for a fixed
algebra $\mathbf{g}$, i.e. the morphisms of the subcategory $\aleph _{
\mathbf{g}}$, we have now to study the morphisms of $\aleph $ that arise
from the inclusion of a subalgebra $\mathbf{g}^{\prime }$ in a bigger
algebra $\mathbf{g}$.

Namely, if $\mathbf{g}^{\prime }$ is a subalgebra of $\mathbf{g}$, the
inclusion morphism associates functorially with any representation functor
of $\aleph _{\mathbf{g}}$ a representation functor of $\aleph _{\mathbf{g}
^{\prime }}$ that splits by a natural transformation of $\aleph _{\mathbf{g}
^{\prime }}$ into irreducible elements according to well known ``branching
rules''. Furthermore, since this process is functorial, any natural
transformation in $\aleph _{\mathbf{g}}$ induces a natural transformation
between the corresponding elements of $\aleph _{\mathbf{g}^{\prime }}$.

\emph{This means that a type }$\mathbf{g}$\emph{\ elementary particle at
the position }$V$\emph{\ seen at the same position }$V$\emph{\ from the
point of view functor }$Hom_{V,\mathbf{g}^{\prime }}(.,V)$\emph{\
decomposes into type }$\mathbf{g}^{\prime }$\emph{\ particles, and that
physical phenomena relating type }$\mathbf{g}$\emph{\ particles are seen
through the point of view functor }$Hom_{\aleph _{\mathbf{g}^{\prime
}}}(.,\rho _{\mathbf{g}^{\prime },V})$\emph{\ as physical phenomena
relating type }$\mathbf{g}^{\prime }$\emph{\ particles.}

We will refer those natural transformations respectively as \textbf{type V }
and \textbf{type VI}. We will also say that a type V natural transformation
induces a \textbf{splitting effect}, and a type VI induces a \textbf{ghost
effect}.

Those two effects express that the elements and morphisms of a category
associated with an algebra $\mathbf{g}$ do appear from the point of view of
the objects of a category associated with a subalgebra $\mathbf{g}^{\prime }$
, not under their true form but as $\mathbf{g}^{\prime }$-type objects
linked by specific relations that only make sense in $\aleph _{\mathbf{g}^{\prime }}$.
By contrast type $\mathbf{g}^{\prime }$ objects do not exist from the point
of view of a $\mathbf{g}$-type object since $\mathbf{g}$ does not respect
any of its subalgebras.

We will thus also refer to the above two effect as \textbf{the weakest law}. 
\emph{It implies, by opposition to all the contemporary theories, that the
``nature'' and the characteristics of an object as perceived from the point
of view of another object depends on the nature and characteristics of this
second object.}

So, having cut the Gordian knot of the space-time gives us the freedom to
explore the algebraic relations between the objects that are usually simply
set down as existing in a preexisting space-time. In other words, by
restricting the scope of the relativity principle to kinematics, modern
physics has implicitly postulated that the appearance and the nature of the
``objects'', and correlatively of the space-time itself, are independent of
any observation and of any observer.

\emph{With the splitting and ghost effects, the absolute relativity theory
frees physics from this constraint}.

\emph{The consequences of this change can not be overestimated. }Let us
review some of them.

\paragraph{Some consequences of the weakest law}

We have seen above that the ``passing of time'' corresponds to a change of
position in the dual of the Cartan subalgebra of a copy of the realified of $%
\mathbf{sl}_{2}(\mathbb{C})$ that we will associate with the electrons. Thus in
each real Lie algebra such that the principal embedding of $\mathbf{sl}_{2}(%
\mathbb{C})$ may be done with this real form, we will get a copy of this
``passing of time'' consistent with the one the electrons.

This remark will help us to determine what are the real forms of the Lie
algebras that can correspond to particles we perceive as able to share our
proper time.

Another important consequence of the weakest law concerns nucleons.

Protons (associated with $G_{2}$) do exist from the point of view of
electrons ($D_{2}\subset G_{2}$)\emph{\ }that orbit them, but, from the
weakest law, no electron does exist from the point of view of a proton ;
therefore, the electrons around the nucleus may move from one to another
since no one exists from the point of view of any nucleon. By contrast, each
of the quarks ($\mathbf{so}_{8}$ or $D_{4}$) included in a proton or a
neutron ($G_{2}$) does exist from the point of view of this one since $G_{2}$
is a subalgebra of $D_{4}$ ; therefore, the three quarks belonging to a
specific nucleon cannot be mixed up with other ones inside the nucleus.

Another very important consequence of the weakest law concerns the real form%
\footnote{{\small See for instance [A.L.O.-E.B.V.], table 4, pp. 229-231 for
all those real form discussions.}} $E_{II}$ of the algebra $E_{6}$, the
highest real non compact dimensional one that contains in its primitive
roots a copy of the realified of $\mathbf{sl}_{2}(\mathbb{C})$ that represents
electrons. Its real rank is $4$, and, since its Dynkin diagram has an $S_{2}$%
-symmetry, it has a singular subalgebra defined by folding, namely the split
form $F_{I}$ of the algebra $F_{4}$, the real rank of which is also $4$

Now, the complex algebra $F_{4}$ contains the direct sum algebra $%
A_{1}\oplus G_{2}$. This means that any representation of $E_{6}$ (or $F_{4}$%
) may be seen as a representation of $A_{1}\oplus G_{2}$, a direct sum that
corresponds to the couple lepton-nucleon. Thus, the atoms we are made of,
will appear as representations of this direct sum algebra, and instead of
perceiving $E_{6}$-bosons as they are, and consequently as massless like any
boson, we perceive them as split in an $A_{1}$ type (anti-) particle and a $%
G_{2}$ type particle\footnote{{\small Let us notice that }$su_{2}${\small \
(resp. }$su_{1})$ {\small is the real compact form of the 2-dimensional
(resp. 1-dimensional) linear group that acts on any Cartan subalgebra of }$%
g_{2}${\small \ (resp. }$su_{2}${\small ) explains why the gauge invariance
group associated with the weak interaction is usually represented by the
product }$SU_{1}\times SU_{2}${\small .}}.

Considered separately, $G_{2}$-bosons, since they are not the true ones for
the description of the weak interaction, may naturally appear as having a
mass exactly in the same way as a $1$-dimensional observer with our time
direction would perceive a photon as a very short-lived massive particle.
This approach gives a way to have an order of magnitude of the masses of W-
and Z-bosons, and therefore to explain their surprisingly high values.

At a smaller scale, the inclusion of algebras $D_{4}\subset B_{4}\subset
F_{4}$ means that $E_{6}$-bosons may also be seen as $D_{4}$%
-representations, which means that one can also interpret their action at
the level of quarks.

Let us note furthermore that, if, as above guessed, the weak interaction
as a true physical phenomenon corresponds to the type II isomorphism
associated with $E_{6}$, seen with some ghost effects from our point of
view, the space-time from the point of view of the observers we are, made of
both neutrons and protons, containing both electrons and neutrinos, should
correspond to the Cartan subalgebras of two non connected copies of those $%
E_{6}$ algebras that for some phenomena can be grasped globally through
their folding in $F_{4}$.

But, since\ the algebra $F_{4}$\ is fundamentally asymmetric, the space-time
associated with it in the above sense does not preserve the left-right
symmetry (symmetry P) respected by the space-time associated with $D_{4}$.
Thus, the symmetry we assign to the space-time does not come from its true
structure, but also from ghost effects induced by the symmetries of the
particles we are made of\emph{.}

\emph{In other words, if space-time were the one associated with }$D_{4}$%
\emph{\ particles, all the physical phenomena would respect the symmetry
P. The fact that this is not experimentally the case justifies the above
guess (that may also be seen as a mathematical consequence of the absolute
equivalence principle applied to }$E_{6}$ \emph{algebras).}

Finally, let us notice that\emph{\ this dependence of the perceived
objects on the perceiving observer extends also to kinematics effects.}

For instance, we have already mentioned that neutrinos have a null real rank
and thus should have, as we will see, a null mass. But, from the point of
view of an electron, they do appear with a ``relativistic'' mass that is in
fact inherited from the non associativity of the structure of the time of
the electrons (or nucleons) we are made of. This ``relativistic'' mass
allows us to explain oscillations with other flavored neutrinos, therefore
avoiding the contradictions that would arise from the existence of a speed
of light massive particles.

All those examples are hopefully sufficient to give a first idea of the
various physical phenomena explained by the weakest law : we are thus now
ready to go to the next step by examining how those phenomena arise through
the point of view functor.

\paragraph{The impact of the point of view functor}

Any embedding of $\mathbf{g}^{\prime }$ into $\mathbf{g}$ also induces also
a surjective homomorphism from the dual algebra $F_{h}(\mathbf{G}^{\prime })$
into the dual algebra $F_{h}(\mathbf{G})$ that allows us to associate with
any representation of the first a representation of the second.

Let us note that in the non quantized case, such dual algebras $F(\mathbf{G%
})$ are algebras of functions, and the functor is a trivial restriction
functor. Furthermore, since all those algebras are commutative, their
irreducible representations are one dimensional and thus isomorphic. Only
their own dual is non trivial and corresponds to the group $\mathbf{G}$
itself through the Tanaka-Krein duality\footnote{{\small Let us emphasize
here that the Lie algebra being defined on the tangent space at the identity
element of a Lie group, the non quantized figure is not stable under the
adjoint action of the group G on itself.}
\par
{\small Since the adjoint action defines natural transformations, beginning
with the non quantized case would have leaded to global consistency
difficulties.}
\par
{\small That is the reason why we have always considered the quantized case
instead of beginning with the simpler non quantized one. See [V.C.-A.P.]
pp.182-183 for a way (due to Reshetikhin) that precisely builds the
quantization from the adjoint action of the group through an appropriate use
of the Baker-Campbell-Hausdorff formula.}}.

In the quantized case, there is also a family of equivalent one dimensional
representations parametrized by the elements of the Cartan subalgebra (or
the maximal torus in the real compact case), and the weakest law in that
case defines what we propose to call\textbf{\ a projection effect} : any
element of the Cartan subalgebra (or maximal torus) of $\mathbf{g}$ can be
projected onto the one of $\mathbf{g}^{\prime }$ (that is included into),
and defines so a one dimensional representation of $F_{h}(\mathbf{G})$. 
\emph{This means that the process can not be reversed, which justifies the
name of weakest law we proposed before}.

But, the most important fact comes from the case $\mathbf{g}=\mathbf{sl}_{2}$
with $\mathbf{g}$ any Lie algebra, which defines a family of infinite
dimensional representations of$\ F_{h}(\mathbf{G})$ on $l^{2}(\mathbb{N})$ with
only a finite number of terms different from zero, itself isomorphic to the
polynomial algebra $\mathbb{C}[X]$ (which is isomorphic to the ring of
representations of $U(\mathbf{sl}_{2})$)\footnote{{\small See for instance
[V.C.-A.P.] pp. 234-238 and pp. 435-439 for the compact case.}}.

In the real compact case, those representations are parametrized by the
elements $w$ of the Weyl group $W_{\mathbf{g}}$ and those $t$ of the maximal
torus $T_{\mathbf{g}}$ of $\mathbf{g}^{c}$\footnote{{\small See for instance
[V.C.-A.P.] pp. 433-439.}}. We will denote by $\rho _{w,t}$ any such
representation. Those representations will play a fundamental role to
explain ``the passing of time'' and to describe the structure of space-time,
as we will see in the next subsections.

We will call this fundamental consequence of the weakest law the \textbf{%
structural law}. \emph{It will appear as the mathematical foundation of 
\textbf{the fundamental theorem of dynamics}.}

\subsection{Recovering space-time and Lagrangians}

\subsubsection{The paradox of space-time}

Since we just have seen that space-time appears only as a by-product of the
particles on a type by type and relativistic way, \emph{we have now to
explain why its use in physics has been so fruitful for more than four
centuries}.

This apparent paradox comes from two mathematical facts :

\begin{itemize}
\item  The first one is well known : we have just seen that space-time
associated with $\mathbf{g}$ type particles should be defined from its
Cartan subalgebra $\mathbf{h}$ and its dual (the maximal torus $\mathbf{T}$
of the group $\mathbf{G}^{c}$ and its dual in the real compact case). This
Cartan subalgebra is commutative and may thus be seen as the product $\mathbb{C}%
^{r}$ of $r$ copies of $\mathbb{C}$ ($(\mathbb{R}/\mathbb{Z)}^{r}$ in the real
compact case) with $r=\dim (\mathbf{h})$. Since it is a solvable Lie
algebra, all its representations are one-dimensional and its only
interesting structure is the one of a vector space.

But, \emph{despite its triviality, a Cartan subalgebra }$\mathbf{h}$%
\emph{\ encodes almost everything that concerns a complex simple Lie
algebra }$\mathbf{g}$. Namely, the Weyl group that completely characterizes $%
\mathbf{g}$ may be defined from its restriction to $\mathbf{h}$. The root
system, the embeddings of subalgebras can also (up to isomorphisms) be
defined on $\mathbf{h}$ and its dual $\mathbf{h}^{*}$. Furthermore formal
characters that characterize all the representations are also defined on the
Cartan subalgebra. In the real compact case, the character functions,
defined on the maximal torus $\mathbf{T}$, generate a topologically dense
subset of the class functions, and every element of $\mathbf{G}^{c}$ is
exactly conjugate to $\mid W\mid $ elements of $\mathbf{T}$. The system of
primitive and fundamental roots which permutations or embeddings in the
roots system define the above type II and type III natural transformations
correspond clearly to Cartan subalgebras (or maximal torus) properties.

Turning to the real non compact algebras, it is also well known that all the
real forms are characterized by a Satake diagram, itself also connected to
the diagram of primitive roots included in $\mathbf{h}^{*}$. Finally, all
the finite dimensional representations of $U(\mathbf{g)}$ and its
deformations $U_{h}(\mathbf{g)}$ are indexed by the weights lattice that is
included in $\mathbf{h}^{*}$, while, as aforesaid, the representations of
the dual quantized algebra $F_{h}(\mathbf{G})$ are indexed\footnote{{\small 
See for instance [A.C.-V.P]. Th.eorem 13.1.9, page 438.}} by the product of
the Weyl group and the Cartan subalgebra $\mathbf{h}$ itself (or the maximal
torus $\mathbf{T}$ in the real compact case).

It is thus understandable that, \emph{by putting everything ``by hand'' on
the vector spaces that correspond to Cartan subalgebras, physics has been
able to induce directly from experimental facts many properties that
characterize Cartan subalgebras of specific simple Lie algebras, although,
from a mathematical point of view, those properties could have been directly
deduced from the analysis of each Lie algebra.\emph{\ }}

The usual inductive method of physics has also been made easier by \emph{
the following, more subtle, mathematical fact that stands at the origin of
Lagrangians and of their universal application}.

\item  The Lie algebras $\mathbf{sl}_{n+1}$ of type $A_{n}$ with $(n>1)$
that act canonically on vector spaces of dimension $n+1$ without any
specific structure have a very unique property\footnote{{\small This fact is
easy to check directly by using plethysm. See also [A.C.-V.P.], pp.387 for
a reference.}}. Namely, from the Jacobi identity, the tensor product $
\mathbf{sl}_{n+1}\mathbf{\otimes sl}_{n+1}$ contains, as the product $
\mathbf{g}\otimes \mathbf{g}$ for any other Lie algebra $\mathbf{g}$, a copy
of the adjoint representation. \emph{But in the }$A_{n}$\emph{\ case, it
contains also a second copy of }$\mathbf{sl}_{n+1}$\emph{\ that belongs to
the symmetric part of }$\mathbf{sl}_{n+1}\mathbf{\otimes sl_{n+1}}$. The
projection $\pi $ on this second copy is given by\emph{: }$\pi (x\otimes
y)=xy+yx-\frac{2}{n+1}trace(xy).Id$.

Any element of $Aut(\mathbf{sl_{n+1}})$\textbf{\ }\footnote{considered as the tensorial product $sl_{n+1}\otimes sl_{n+1}^{\ast}$} corresponds to an element
of $\mathbf{sl_{n+1}\otimes sl_{n+1}}$ through the Killing form, and,
therefore, has a projection through $\pi $ onto a symmetric representation
of $\mathbf{sl}_{n+1}$. This projection is clearly null for the image of the
identity map that corresponds to the Casimir operator, but it can be non
trivial in other cases. Thus, we may associate a bilinear symmetric form on $
\mathbf{sl_{r}}^{*}$ that is a representation of $\mathbf{sl}_{r}$ with any
Cartan subalgebra of rank $r$ with $r>2$. This correspond to the Lagrangian
that we have defined and used in our first section.

Let us notice finally that for algebras with $r=2$, the above component of $
\mathbf{sl_{r}\otimes sl_{r}}$ is null, and the Lagrangian is reduced to its
free part as we shall see below for the electromagnetic interaction (that
will appear as a type II interaction that changes the compact real form $
\mathbf{so}_{4}$ to the realified $\mathbf{so}_{1,3}$\textbf{\ }of $\mathbf{
sl}_{2}(\mathbf{\mathbb{C}})$).
\end{itemize}

\subsubsection{The structure of space-time}

Let us come back before to space-time as it appears now associated with each
type of Lie algebra and therefore to each type of particles.

Let us keep the above notations and define \textbf{space-time associated
with }$\mathbf{g}$\textbf{\ }(or $\mathbf{g}$\textbf{-space-time)} as the $r$
-dimensional vector space as above defined by the dual $\mathbf{h}^{*}$ of a
Cartan subalgebra $\mathbf{h}$ supposed chosen in the beginning of the
construction.

The first two properties of this space-time are that it is canonically
equipped with the bilinear symmetric form defined by the restriction to $
\mathbf{h}$ of the Killing form $kill_{\mathbf{g}}$, and that it contains
the lattice of maximal weights that indexes all the finite dimensional
representations of $U_{h}(\mathbf{g})$.

\emph{Since this lattice is defined as a copy of }$(\mathbb{Z)}^{r}$\emph{,
it is clearly not invariant through }$e^{i\phi }$ \emph{changes of phase :
space-time arise first as a discrete real mathematical object. This leads us
to work on real Lie algebras without any assumption of an ``external''
breach of symmetry in }$\mathbb{C}$\emph{.}

Now\footnote{{\small We follow here [A.L.O.-EB.B.V.], page 133.}}, any real
form is defined by an invariant conjugate linear form $\sigma $. Two linear
forms $\sigma $ and $\tau $ are said to be \textbf{compatible} if they
commute, which means that the composite map $\theta =\sigma \tau $, that is
a linear one, is involutive. Any real form is compatible with a real compact
form. We may thus begin by working on the compact real form $\mathbf{g}^{c}$
of $\mathbf{g}$, and go then to any other real form by applying an invariant
involutive\emph{\ complex} transformation\footnote{{\small It is
interesting to notice that there always exist on the compact form a Haar
measure that authorizes convergent integral calculus. The pull back of this
calculus on a non compact real form defines canonically a way to make
convergent calculus on this non compact real form that is generally itself
equipped with a divergent invariant measure. This remark could explain in
our context the success of renormalizations methods of QFT, but we did not
go deeper into this way.}}.

Let us emphasize that, since such a transformation corresponds to a type II
or type III natural transformation, \emph{the particles associated with
two compatible real forms do coexist in space-time, and are related by a
true physical phenomenon}.

For instance, neutrinos do coexist with electrons and positrons\footnote{
{\small This fact can be confirmed by a well known fact on Lie groups. The group SO4 splits into two copies of the group SO3. Since
the multiplication by i in C acts as an orthogonal rotation on the real
plan, the Weyl symmetry may be seen as applying such a rotation on the real
plan, which corresponds exactly to the Lie bracket [i,j] in the su2 algebra.}
\par
{\small Applied on any two directions, this means that applying the above
involution is equivalent to go from the (-) connection on the sphere that is
the usual one without torsion, to the (+) connection that has a torsion
defined by the Lie bracket on the tangent space at the neutral element.}
\par
{\small The compact form then appears as the torsion -free manifold
associated with this second one.}
\par
{\small There is another possibility to define the (+) connection by
left-right conjugation, while the compact form is unique. See for details
[S.H.], pp. 102-104.}} as associated respectively with the compact forms $
\mathbf{so}_{4}$ and $\mathbf{so}_{1,3}$ of the real algebra $\mathbf{sl}
_{2}(\mathbb{C})$. Furthermore, as associated with the same compact form,
neutrinos and antineutrinos should be the same particle, but we will see
that they differ slightly due to dynamical ghost effects that generate their
mass\footnote{{\small This mass may be increased by type III transformations
that induce mu and tau neutrinos.}} as we will explain in the next
subsection. In the same way, dark quarks should coexist with usual ones
corresponding with the compact form $\mathbf{so}_{8}^{c}$ and be folded in
dark neutrons associated with the compact algebra $\mathbf{g}_{2}^{c}$.

\emph{On the other hand, if two non compact forms are associated
respectively to the involution }$\theta $\emph{\ and }$\theta ^{\prime }$
\emph{\ that do not commute, which is the general case, the corresponding }
$\tau $\emph{\ and }$\tau ^{\prime }$\emph{\ do not commute and the two
real forms are not compatible : the corresponding particles can not coexist,
and therefore, in general only one real form of each type of algebra can
coexist with the compact form. This remark has been used extensively in our
section II, since we will have to find only one non compact real form for
each type of particle\footnote{{\small This uniqueness of the non compact
real form does not implies that the corresponding particle appears only
under one form : indeed, as we have seen before, there can be relativistic
type III effects that change the ``flavor'' of the particles, and also
purely dynamical ones, as we will explain below. }}.}

We have now to analyze how $\mathbf{g}^{c}$-space-time is seen through the
point of view functor. Going to the corresponding analysis for other
algebras will then be easy by applying the transpose of the above
involution, which is legitimate since, by tensoring by $\mathbb{C}$, we can
take everything on the complex field, and then apply the above complex
involution.

Now, beginning with the $\mathbf{su}_{2}$ case, it is known\footnote{{\small 
See for instance [V.C.-A..P.], page 437.}} that the representations of $
F_{\varepsilon }(\mathbf{SU}_{2})$ are parametrized by the maximal torus $t$
with $t\in \mathbb{C}$, $\left| t\right| =1$, and are either one-dimensional $
\tau _{t}$ or infinite dimensional $\pi _{t}$ given by, for linear forms $
a,b,c,d$ arranged as $\left( 
\begin{array}{ll}
a & b \\ 
c & d
\end{array}
\right) $ in the usual presentation :

\begin{center}
$\tau _{t}(a)=t$ ; $\tau _{t}(b)=0$ ;$\tau _{t}(c)=0$ ; $\tau _{t}(d)=t^{-1}$
;

$\pi _{t}(a)(e_{k})=(1-\varepsilon ^{-2k})^{1/2}e_{k-1}$ ; $\pi
_{t}(b)(e_{k})=-\varepsilon ^{-k-1}t^{-1}e_{k}$ ; $\pi
_{t}(c)(e_{k})\varepsilon ^{-k}te_{k}$ ; $\pi _{t}(d)(e_{k})=(1-\varepsilon
^{-2k-2})^{1/2}e_{k+1}$,
\end{center}

with the $e_{k}$ belonging to the formal one variable real polynomial
algebra.

The first one may be connected with zero dimensional symplectic leaves of
the Poisson Lie group $\mathbf{SU}_{2}$, while the second corresponds to the
two-dimensional ones (associated with the opposite face of the sphere $S^{2}$%
, and therefore to the application of the Weyl symmetry).

Finally we see that, in that case, the representation of $U_{\varepsilon }(%
\mathbf{su}_{2})$ are parametrized by a lattice in the dual of the maximal
torus, and that representations of the dual Hopf algebra $F_{\varepsilon }(%
\mathbf{SU}_{2})$ are parametrized by the maximal torus itself. Furthermore,
if, by putting $X=(x+\frac{1}{x})$, we identify the above polynomial algebra
with the algebra of characters, and correlatively of representations, of$\
U_{\varepsilon }(\mathbf{su}_{2}),$ we see that successive applications of
the diagonal elements of the dual Hopf algebra induce a move of increasing
degrees representations which we will refer to as \textbf{a growth of
representations} and that may be seen as corresponding to successive
applications of the point of view functor : this will be the origin of the
passing of time we will describe in the next subsection. Reciprocally, a
diagonal element of the universal enveloping algebra generates a
one-parameter group that induces a continuous move on the representations of 
$F_{\varepsilon }(\mathbf{SU}_{2})$ as parametrized by $t$.

\emph{We have thus a perfectly reciprocal situation. It is well-known that
particles mysteriously have a double nature, corpuscular and undulatory ;
the structure of the time that arises canonically as we just have seen shows
that in some sense, time itself has a very precise double structure,
spectral and undulatory, but there is nothing mysterious in this situation :
the one corresponds to the Quantum Universal Algebra ; the second to its
(restricted) dual as it appears after application of the point of view
functor}.

Such a structure that arises canonically in our context was probably
difficult\emph{\ a priori} to postulate, although it is not so far from
many recent approaches.

As mentioned before, this situation may be generalized to any compact algebra $
\mathbf{g}^{c}$ by using the principal embedding of $\mathbf{su}_{2}$ in the
Weyl chamber of $\mathbf{g}^{c}$, and then composing with applications of
the Weyl group : this gives the aforesaid $\rho _{w,t}$ representations.
\footnote{{\small See ibidem pp. 437-439.}}.

\emph{We therefore have exactly the same situation for any real compact
algebra }$\mathbf{g}^{c}$\emph{\ as the one we have just described for }$
\mathbf{su}_{2}$\emph{, except that we have now an }$r$\emph{
-dimensional torus, and that there is no more one copy of the algebra of
representations, but as many as there are elements of the Weyl group }$W$
\emph{, each copy being connected to the others by a type III
transformation.}

\emph{Applied to }$\mathbf{g}_{2}$\emph{\ (or }$\mathbf{so}_{8}$\emph{
), this situation gives rise to the ``flavors'' of particles: their proper
time may stand, up to a generalized Lorentz transformation, in any of the
copies of the Weyl chamber of }$\mathbf{g}_{2}$\emph{\ that corresponds to
the nucleons we are made of. Furthermore, since generalized Lorentz
transformations can partially cover two distinct copies of the Weyl chamber
of }$\mathbf{g}$\emph{, oscillations of ``flavors'' can arise.}

\emph{The structure of space-time is finally totally defined by the
characteristics of the algebra it is generated by : it nevertheless always
has a quite sophisticated fibered continuous structure combined with a
discrete one, each one of those two components acting on the second.}

\section{The Mass Quantification Theory}

As announced at the twelfth step of section 1.4, the general methodology that we will use will take as a reference for our time the time of the electron. Then, we will do the calculations in this time frame considered as embedded in the algebra that caracterizes the type of particle chosen, alongside the direction defined by the principal embedding of $sl_{2}$, sometimes composed with the appropriate transformation of the Weyl group of the Cartan subalgebra.

The principal embedding and the Weyl group are indeed the two elements that define at each point of the compact representation associated with the particle considered the creation / annihilation operators which generate the time of this particle. The MQT therefore implies a preliminary step that will make more precise our vision of the electron.

It will be the first point of this section, and the second one the calculation of the Newton ``constant'' and the third point the application of the genreral guidelines  defined in the previous sections from a frame tied to an electron as an an bserver to some examples that will illustrate the possibilities offered by the MQT.

\bigskip

\emph{The computations given as examples below, by no means extensive, have the main goal to show that the ART provides numerous numerical results that can be checked by measures, and that the MQT can appear as a new branch of physics. The authors recognize that these calculations have to be upgraded and extensively completed for other types of particle}

\subsection{Calculations for the electron}

The electron is defined from the realification of the complex algebra $SL_{2}(\mathbb{C})$. The tensorial product by the adjoint representation will amount to a displacement of two units in the dual of the Cartan subalgebra. On the other hand, it is also on the adjoint representation of $SU_{2}$ (which is like the sphere $SO_{3}$) a rotation of $\pi$ angle on the maximal torus. 

A necessary condition of coherence for the fact that the KZ-monodromy that defines the passing of time is not changed by the rotation on the torus on which the representations of $F_{h}(SL_{2}(\mathbb{C}))$
are indexed, is that the quadratic form which connects the Cartan subalgebra to its dual be such that to 2 on the dual corresponds $\pi$ on the torus.
 
The time unit for the time of the electron corresponds therefore to 4 units on the dual (this corresponds to two successive monodromy in the adjoint algebra, each one of them inverting the algebra orientation because of the flip operator) or to a translation of 4 in terms of the Yangians.
 
Thus a creation cycle of one unit of time for the electron is like a $2\,\pi$ rotation on the maximal torus.

The involution that generates the real form of $Sl_{2}(\mathbb{C})$ from the compact form of $SO_{4}(\mathbb{C})$ being also the transformation that exchanges algebraically the left with the right, one period of time of the electron will then be like $2\,\pi$ on pure imaginary direction.

Now, as we have done before, let us represent the Cartan subalgebra in the 0 and 1 directions (or the real and imaginary ones) associated to the space-time of the electron, which corresponds to $su_{1,1}$. The dimension 2 and 3 will correspond to the compact part, and therefore will be be associated with its charge. The ``hyperbolic'' rotation in $su_{1,1}$ is the image of the rotation in $su_{2}$ before the application of the involution which transforms the compact form into the real one. 

The ratio between these two rotations is therefore a characteristic of the relative energies of these two components (electromagnetism and mass). It is thus given by the fine structure constant $\alpha$. And we just showed that we have to use instead  $\alpha'=\frac{\alpha}{2\,\pi}$, that characterizes the ratio between the energy of the electron that comes form his mass and the electromagnetic energy, because of our normalisation. 

This modelisation allows several calculations.
The figure before the application of the Weyl transformation contains a right-left symmetry if we read the product $e_{0}\wedge e_{1}$ as the dual of $e_{2}\wedge e_{3}$ (Hodge-$\ast$ operator).
\begin{enumerate}

\item The rotation that happens on directions 0 and 1 before the application of the involution induces an equivalent rotation on directions 2 and 3 because of the Hodge-$\ast$ operator. 
\item The rotation in the 2 and 3 directions that corresponds to the electron charge is thus changed by a precession coming from the previously described rotation that comes from the passing of time of the electron. As a first approximation, it is then $$p=1-\frac{1}{1+\alpha'}$$

To be more precise, the measure to consider on the maximal torus has to take into account the fact that this torus is in reality a main circle of a sphere of radius 1 that stands for $SO_{3}$. The norm to take into account at the second order is therefore $1-\frac{1}{3}\alpha'^2$ (the limited development at the second order of the Riemannian form in the neighborhood of a point), and thus we obtain $$p=\alpha'-\alpha'^2-\frac{1}{3}\alpha'^2$$

If we compare this result with the coefficients that comes from the QFT, usually expressed as a function of $\alpha''=\frac{\alpha}{\pi}$, and that are $\frac{1}{2}\alpha''-0.3285\alpha''^2$, we see that the difference at the second order is, taking into account the measured value $\alpha'=0.00116140981411$ $$(\frac{1}{3}-0.3285)\times\alpha''^2\cong 6.6057\times 10^{-10}$$

The QFT includes higher components that come either from the electromagnetic part or from potential other particles. We did not try in this first paper to represent this phenomena. Roughly, our first approximation gives $$p_{calc}\cong 0.001159611317$$ compare to the measured value $p_{exp}\cong 0.001159652$

In our calculation we did not take into account the asymmetry of time which comes from the passage from the KZ-algebra to the DZ-algebra. We can guess that at least a part of the difference comes from this approximation.

\item On the other hand the fact that the left-right conjugation exchanges the directions 0 and 1 with the directions 2 and 3 implies that the precession $p$ is the right equivalent of $\alpha$, the fine structure constant. Each term is calculated by supposing the other fixed. We will then introduce the coefficient $$\psi=\frac{\alpha'}{p}$$ to represent the impact of the left right conjugation on the electron representation. 

\item We will calculate now the position $N$ of the electron in its own mobile frame. By reasoning in the ``ghost time'', we consider that the electron in the past is like it is now. So by getting $N$ we will get the age of the universe.

We have seen that a move of two units in the dual of the Cartan subalgebra is equivalent because of the Casimir coefficient to a variation of $\frac{1}{\sqrt{C_{2}(n)}}$ in the subalgebra itself (cf. the previous sections I and II). This reasoning applies to the electron and its associated bosons (that is to say to all the successive value of $n$), and therefore we have to consider that the variation of the quadratic form on the Cartan subalgebra is given by the sum 
$$\Sigma _{N}=\sum\limits_{k=1}^{N}\frac{1}{\sqrt{(k+1)^{2}-1}}$$

\emph{This length is measured in the basic algebra, and thus in the dual
algebra of the one the time belongs to. Since the energy is given by $\frac{\partial\mathcal{L}}{\partial\dot{z}}$ with $\mathcal{L}$ the above defined Lagrangian, it is merely proportional to the quantity $\dot{z}$ in the dual of time. Therefore, it is
thus natural to define its inverse as \textbf{the energy associated to the
position} $N$.}

Since even and odd representations correspond to distinct groups
representations, and thus to distinct quantum groups, it will be more convenient
to apply the above construction separately to even and odd representations
that will correspond to distinct particles. One get then two distinct
summations :

$S_{N}=\sum\limits_{k=1}^{N}\frac{2}{\sqrt{(2k-1+1)^{2}-1}}$ and $
T_{N}=\sum\limits_{k=1}^{N}\frac{2}{\sqrt{(2k+1)^{2}-1}}$.

A small computation shows that for $N>10^{6}$, the sum $\Sigma
_{N}=S_{N}+T_{N}$ may be approximated (with a less than $10^{-11}$
precision) by :

$\Sigma _{N}=\ln (2N)+0.08396412352$.

From the modelling of the electron in the MQT previously described
we have seen that the fine structure constant $\alpha$
corresponds to the ratio between the electromagnetic energy (that does not
varies with time) and the energy corresponding to the inertial mass (that
decreases with the above law) of the electron. It can thus be used to
compute $2N$ with the above formula.

More precisely, we will get from hyperbolic trigonometric considerations\footnote{This comes directly from the exchange of the direction 0 and 1 (that is direction of the real axis and the direction of the imaginary one) after we take into account the invariant quadratic form and the passage from an euclidean form to a lorentzian one}
that we must have 

$$\Sigma _{N}=\frac{2}{\pi }\sqrt{\frac{1}{\alpha ^{2}}-1}$$ and thus 

$$2N=\exp(\frac{2}{\pi }\sqrt{\frac{1}{\alpha ^{2}}-1}-0.08396412352)$$

We will state again $\alpha ^{\prime }=\frac{\alpha }{2\pi }$ and the
measured value of $$\alpha ^{\prime }\cong0.00116140981411$$
gives 

$$2N=7.08432804709\times 10^{37}$$

To come back to the international system of units, we use the numerical value of the
speed of light $c=2997932458\ m/s$ to define the unity of length and Planck
constant $h=6.6260755\times 10^{-34}\ kg.m^{2}/s$ to define the unity of mass.
We get $\frac{h}{c^{2}}=7.37250327\times 10^{-51}kg.s$.

With the mass of the electron $m_{e}=9.109384\times 10^{-31}\ kg$, this leads
to the period of the electron 

$$T_{e}=8.09355159\times 10^{-21}\ s$$

that defines the apparent age (or ``ghost'' age) of the universe from our point of view as 

$$(2N)T_{e}=5.73373745\times 10^{17}s\approx 18,18\times 10^{9}$$ in years\footnote{This age is of the same order of magnitude as the one
actually recognized by contemporary physics. It is nevertheless slighty
higher, but the absolute relativity theory leads to a view on the
cosmological beginning of the universethat goes far more slower scenario
than the usual ones.
\par
But the most significant confirmation of this value will be the
computation below of the Newton constant that results from the absolute
relativity theory that is based on this value and fits perfectly with
experimental values.}.

\end{enumerate} 

\subsection{The Newton constant calculation}

The ART makes of the gravitation a consequence of the passing of time. It should then be possible to calculate the value of the Newton ``constant'' (remember it appears such from the present of an observer looking at his ``ghost'' past) from the previously calculated value of $N$.

The variation of the measurement unit in the direction of time corresponds to the component of the trace operator projected on the identity (cf. the coefficient $\frac{\tau}{2}$). A given variation in 2 dimensions will therefore have an spatial impact twice the one in 4 dimensions. This  is equivalent to replace $N$ by $\frac{N}{2}$. 

 So to find the electron we have to change the coupling coefficient by two. 

Instead of calculating the Newton constant by itself, we will instead compare two dimensionless terms that should be equal if the intrinsic coupling coefficient is indeed 1 as the ART predicts. Such a checking is obviously the same as the calculation of the constant, depending only on the choice of units.

The right term  is given by the effects of the two coefficients which after being multiplied make the volume form changes (here the volume form is a surface form for the space-time of the electron). If we take into account the 2 coefficient afore mentioned, we obtained, with $\alpha\cong 2\pi\times0.00116140981411$ and the previously computed age of the universe $N\cong 7.08432804709\times 10^{37}$:
$$\frac{\omega_{t}}{\omega_{0}}=2\times\frac{\alpha}{\sqrt{1-\alpha^2}}\times \frac{1}{N}=2.06019465385\times 10^{40}$$

On the other side, if $m_{e}$ is the mass of the electron, and if we take for $G$ the value measured as of today, that is $6.67259\times 10^{-11}$ in the SI unit system, and if we do not forget that the initial mass to take into account is modified by the duality, that is $(\psi\alpha')^{-1}$ instead of $\alpha'^{-1}$, with $\alpha'=\frac{\alpha}{2\pi}$, the volume form should be modified by the dimensionless coefficient defined by :$$\frac{\omega_{t}}{\omega_{0}}=\frac{G(\psi\alpha'^{-1}m)^2}{c\,h}$$ which is $$\frac{6.67259\times 10^{-11}\times(\frac{1}{1.0015156511\times 0.00116140981411}\times 9.109384\times 10^{-31})^2}{299792458\times 6.626176\times 10^{-34}}$$ which is $$\cong 2.06016657202\times 10^{-40}$$

The ratio between the two coefficients is therefore $$\frac{2.06019465385\times 10^{40}}{2.06016657202\times 10^{-40}}\cong 1.00001363085$$ so the difference is less than 0.002 \%, which corresponds to the margin of error on the numerical coefficients used and to some evolutions of the measured physical data, and also comes from the asymmetry of time ...

We have therefore confirmed that $$G=\frac{\alpha_{3}\,c\,h}{(2\,\pi\psi m)^2N\sqrt{1-\alpha^2}}$$

The quality and precision of this result confirms at the same time the generalised Einstein equation we have established (and especially its dependance of the dimension of the space time associated with the particle) and also our calculation of the age of the universe, especially because it involves the exponential of a number around 100 : the precision we have obtained around 0.001 \% show that the precision was before the application of this exponential around $10^{-7}$.

\subsection{Mass calculations of the proton and the neutron}

Having done the calculations in the case of the electron, now we only have to follow the guidelines given in the previous parts of this text.
The proton and neutron will then be of type $G_{2}$, and we will look first at the particle without an electric charge.

We will work in the normal (split) real form $g$ of the the complex lie algebra $g_{2}$ of the complex Lie group $G_{2}$.

We use the usual euclidean representation of the roots of $g_{2}$, with the longest one having a length of $\sqrt{2}$.

In that case, the first simple root is $\alpha_{1}=(-1/3,2/3,-1/3)$ and the second simple root $\alpha_{2}=(1,-1,0)$. The weight of the adjoint representation is the longest fundamental weight $$\alpha_{6}=3\,\alpha_{1}+2\,\alpha_{2}=(1,0,-1)$$
The principal embedding of $sl_{2}(\mathbb{C})$ in $g_{2}$ has the following representation $$f=18\,\alpha_{1}+10\,\alpha_{}{2}=(4,2,-6)$$ and a norm of $56$.

The cosinus of the principal embedding angle $a$ with the adjoint representation root is then
$$\cos{a} = \frac{(\alpha_{6},f)}{\sqrt{(\alpha_{6},\alpha_{6})(f,f)}}=\frac{5}{2\sqrt{7}}$$

Because we want to compare the time of the electron with a particle with a time lagged, we have to work in the dual of its Cartan subalgebra, that is the one after the weyl symmetry is applied. It is the algebra of the coroots, in which the fundamental embedding has the representation $5\,\alpha_{1}+3\,\alpha_{2}$ (the half sum of the positive roots) after the appropriate exchange and renormalization of the two simple roots. In that case, with the same convention for the norm (the longest root has a $\sqrt{2}$ length) its norm is $\frac{56}{3}$, that is a third of the norm of the initial principal embedding vector before the transformation.

\bigskip

Therefore, the Dynkin index of the principal embedding being $28$, we have to divide it by $3$ when we work in the dual because of the time lag. By taking into account the standard representation of the electron we begin with, we also have to correct it by the ratio of the Casimir coefficients determining the passage from the adjoint to the standard representation in $sl_{2}(\mathbb{C})$, that is $\frac{8}{3}$.

Finally, the ratio of the intrinsic forms we have to consider is $$i=28 \times \frac{1}{3} \times \frac{8}{3}=\frac{224}{9}$$

Now we can compute the lag in the direction of the principal embedding of the time of a particle of type $G_{2}$ compared to the one of the electron. This ``lag'' will allow us to get the $\cosh{t}$ that has to be used, and then we will get the mass of the particle considered.

\bigskip

In the mobile frame of the electron, the change in the coefficient is therefore given by $$t=(1-\cos{a})\times 2\frac{\sqrt{1-\alpha^2}}{\alpha}\times \frac{2}{\pi} \times \frac{1}{i}\cong 0.386181207486$$

The appearance of the coefficient $\frac{2}{\pi}$ was explained earlier. Taken into account the precession, we find for this particle, that we call the neutron, the following mass $m_{n}$, if $m_{e}=9.10938215(45).10^{-31}\ kg$ is the mass of the electron $$m_{n}=2\cosh{t}\times\frac{m_{e}}{\alpha'(1+\alpha)}\cong1.67488836.10^{-27}\ kg$$ to be compared with the measured mass $m_{n}'=1.67492729(28).10^{-27} kg$, giving a ratio of $\frac{m'_{n}}{m_{n}}\cong1.0002$, next to 0.002 \% of error.

\bigskip

If we calculate now on the other face of $SO_{4}{(\mathbb{C})}$, using a rotation of $\frac{\pi}{2}$ compare to the initial direction of the electron, we should see a particle with the appearance of an electric charge opposite to the one of the electron, and for the calculation we have to use the coefficient $\psi\cong 1.0015156511$ defined and calculated earlier, replacing $\alpha$ by $\frac{\alpha}{\psi}$ everywhere it appears.

So the new  ``lag'' $t'$ becomes $$t'=(1-\cos{a})\times 2\frac{\psi\sqrt{1-(\frac{\alpha}{\psi})^2}}{\alpha}\times \frac{2}{\pi} \times \frac{1}{i}\cong 0.38676658329$$ and the mass of the particle that we can call the proton $$m_{p}=2\cosh{t'}\times\frac{m_{e}\psi}{\alpha'(1+\frac{\alpha}{\psi})}\cong 1.672744.10^{-27} kg$$
which has to be compared with the measured value of $m'_{p}=1.672621637(83).10^{-27} kg$, giving a ratio of $\frac{m_{p}}{m'_{p}}\cong 1.00007$, next to 0.0007 \% of error.

\subsection{Mass calculations for the electrons $\tau$ and $\mu$}

Here, the methodology is the same than before, but simpler because we do not have to compare the metric on the embedding with the one of $g_{2}$. We just move the copy of $sl_{2}$ in $g_{2}$ by the Weyl rotation of the $g_{2}$ root diagram of angle $\frac{\pi}{3}$. The first application of this rotation makes us cross the first zone (angle $\frac{\pi}{4}$) of the root diagram of $so_{4}$, which implies to use the dual, and this has the following consequences :
\begin{enumerate}
\item Use the $\cosh^{-1}$ instead of the $\cosh$ to calculate the ``lag'' $t$.
\item Use the factor $\frac{1}{2}$ instead of $2$ in front of the $\cosh{t}$.
\end{enumerate}

Indeed, the $\frac{\pi}{2}$ rotation of the roots of the realified of $sl_{2}$ needed to pull back the Weyl chamber of $sl_{2}$ in a position that contains the $\mu$ direction is equivalent to the application of the Weyl symmetry that corresponds to the change to the dual. 

To determine the position on the embedding of the electron, we just have to project on it with the $\frac{\pi}{3}$ angle. Therefore, $$t=\cos{\frac{\pi}{3}}\times \cosh^{-1}{\frac{1}{\alpha}}\cong 2.8066887241$$

Then we use the electromagnetic mass of the electron, in $GeV/c^2$ units, that is $m_{e}\cong 0.440023544386\ GeV/c^2$ to get the mass $m_{\mu}$ of the electron $\mu$, $$m_{\mu}=m_{e}\times \frac{2}{\cosh{t}}\cong 0.105931407278 \ GeV/c^2$$ to compare with the measured mass of $0.1056\ GeV/c^2$.

\bigskip

For the electron $\tau$, after another application of the Weil rotation of angle $\frac{\pi}{3}$ of the root diagram of $so_{4}$, we again change the second zone of it by going through the angle $\frac{\pi}{2}$, which implies we go again to the dual and have to multiply by $\frac{1}{2}\cosh{t}$ instead of dividing by it, and use also the $\psi$ coefficient.

The ``lag'' becomes $$t'=\cos{\frac{\pi}{3}}\times \cosh^{-1}{\frac{\psi}{\alpha}}\cong 2.80744603698$$ and therefore the mass $m_{\tau}$ of the electron $\tau$ is $$m_{\tau}=m_{e}\times \frac{1}{2}\cosh{\frac{t'}{\psi}}\cong 1.82146\ GeV/c^2$$ to compare with  the measured value of $1.784\ GeV/c^2$.

\section{Appendix I : On the foundations of the Absolute Relativity Theory}

The transcendental conditions referred to in the first section can be
expressed in a formal mathematical way by using the language of categories.

Since it is not the main purpose of the theory, we only outline here shortly
the framework of such a mathematical formulation that has been useful to
made the main choices that lead to the ART.

\subsection{From graphs to categories}

Representing the objects and morphisms used by a theory as points and arrows
of a graph seems to be the most general algebraic way to begin with a
theoretical construction.

It is then natural, as mentioned in section one, to define \textbf{the point
of view of an object }$a$\textbf{\ on an object }$b$ as the arrows that have 
$b$ as domain (or source) and $a$ as codomain (or target). \emph{We will
assume that any so defined point of view is a set}.

If the objects of the graph are distinct, applying this operation on all the
objects of the graph generates sets of arrows that are not intersecting,
which forbids to make any comparisons between the different points of view
of those objects. It is thus impossible to refute the theory by
experimenting contradictions between the different points of view
represented, which could lead to solipsistic approaches.

Furthermore, there could be isolated objects that have no connection with
everyone including themselves, and such objects do not have any point of
view and are not part of anyone. Thus, their withdrawal would not change the
theory that therefore is not well defined.

By contrast, if we assume that the graph is equipped with a law of
composition of arrows, there begins to be a possibility to make tests of
consistency since, if three objects $a,b$ and $c$ are linked by arrows $\rho
_{c,b}$ from $c$ to $b$ and $\rho _{b,a}$ from $b$ to $a$, there is a
necessary relation between the points of view of $a$ and $b$, namely : the
composed arrow $\rho _{b,a}\circ \rho _{c,b}$ has to belong to the set $\hom
(c,a)$ that represents the point of view of $a$ on $c$. This gives a
possibility to test the consistency of the points of view of the object $a$
and $b$ on the object $c$, but does not allow any consistency test of those
points of view on arrows.

We have thus to consider a fourth object $d$ and an arrow $\rho _{d,c}$.
From the point of view of $b$, the arrow $\rho _{c,d}$ associates an arrow
belonging to $\hom (c,b)$ with any arrow $\rho _{d,b}$. The same applies to
the point of view of $a$ on the same arrow $\rho _{d,c}$. If we restrict
ourselves to the arrows from the point of view of $a$ are defined through $b$%
, it is now easy to express the consistency conditions that is :

\begin{center}
$\rho _{d,a}=\rho _{d,b}\circ \rho _{b,a}=(\rho _{d,c}\circ \rho
_{c,b})\circ \rho _{b,a}=\rho _{d,c}\circ (\rho _{c,b}\circ \rho
_{b,a})=\rho _{d,c}\circ \rho _{c,a}$,
\end{center}

which corresponds to the associativity of the composition law.

Therefore, requiring the composition law to be associative is a necessary
and sufficient condition to make consistency tests on both objects and
morphisms.

We will also assume that there is an identity arrow attached to each object
which means that each object has a non empty point of view on itself and that
one of the corresponding arrows acts as an identity for the operation of
composition of arrows. Those conditions are those that define a
metacategory. For technical reasons, we will assume that this metacategory
is in fact a category and that this category is small enough to see each
object as element of a ``big enough'' set.\footnote{{\small This refers to
the notion of universe as described for instance in [S.M.L.], page 12.}}

We will call \textbf{preuniverse} this category that contains those objects
and arrows used to describe the theory, and we will call it $\aleph .$

The above operation of taking the point of view becomes a contravariant
functor that we will call\textbf{\ the point of view functor}. According to
Yoneda Lemma the preuniverse may be embedded in the category of point of
view functors by the operation that associates the functor $\hom (.,a)$ with
any object $a$, and the corresponding natural transformation with the arrows
between any two objects $a$ and $b$. Yoneda Lemma implies that one can hope
to access to the preuniverse through the category of all the point of view
functors with the corresponding natural transformations.

A functor $F$ in a category $C$ will be said to be representable if there
exists some object $a$ belonging to $C$ such that $\hom _{C}(.,a)$ is
naturally equivalent to $F$.

The point of view functor is contravariant which means that it is a
covariant functor from the opposite category that is obtained by reversing
all the arrows.

The next conditions we will require for the theory we are beginning to build
is that it has ``enough'' objects to make usual composed functors
represented : this will mean that the theory is required to be able to
represents by an object the usual composed functors in a way we will precise
now.

\subsection{From categories to additive categories and R - modules}

We follow here [S.M.L.] pp.198-201 and 209.

Since we want to work on set of arrows coming to a point, it is natural to
require that the product of two hom-sets $\hom (b,a)$ and $\hom (c,a)$
defines a new object $d$ \emph{that does not depends on }$a$, which is, as
above, a condition for authorizing comparisons between points of view. This
correspond to the existence of a \emph{coproduct} associated with any pair
of objects. In the same way, by considering the opposite category, we see
that the category need also to contain a\emph{\ product} of any two
objects.

One may ask also that there is an object the point of view of which contains
all the objects of the category : otherwise one can formally add it in the
category. Imposing the same condition to the opposite category leads to
postulate the existence in the category of an initial object. In order to
have some invariance by the ``taking the opposite'' operation, one
postulates that those two objects coincide, which means that the category
has a zero element.

Let us add two other technical conditions, namely :

- every arrow in the category has a kernel and a cokernel ;

- every monic arrow is a kernel, and every epi is a cokernel.

In an informal language, the first of those conditions is again a condition
requiring the category to have enough objects to make the point of view
functor precise enough in order to allow to represent those fundamental
(abstract) characteristics of arrows. The second one asserts that the
category also has enough arrows in order to make possible the reciprocal way.

The key point is that those conditions are sufficient to ensure that the
considered category is an Abelian category.

Finally, an embedding theorem referred to as Lubkin-Haron-Feyd-Mitchell
theorem in the above reference, allows us to consider the category as a
category of $R$-modules for some suitable ring $R$.

We are therefore now working on such a category.

For an aim of simplicity here, we restrict ourselves to categories of $R$%
-modules that are also complex vector spaces.

\subsection{From additive categories to monoidal categories of modules over
quasi-Hopf algebras.}

Since we access everything through the above point of view functor, one can
ask that there should be also objects in the category that represents (as an
appropriate adjoint) the successive applications of this functor.

This leads to consider that we have also to be able to define tensor
products of the $R$-modules we are now working on. This means that $R$ has
to be a quasi-bialgebra. Since we want furthermore to have the possibility
to exchange left and right, we need an antipode operator.

This leads to the idea \emph{to restrict ourselves to the cases where }$R$%
\emph{\ is quasi-Hopf algebra.}
\pagebreak

\section*{References}

\bigskip

[ALO-EBV] Onishchik, A.O.,Vinberg E.B. (Eds.)

Lie Group and Lie Algebras III, Encyclopedia of Mathematical Sciences, Volume 41, Springer

\bigskip
[SML] Mac Lane, S.

Categories for the Working Mathematician, Second Edition, Graduate Texts in Mathematics, Springer

\bigskip

[WF-JH] Fulton, W., Harris, J.

Representation Theory, a First Course, Graduate Texts in Mathematics, Springer

\bigskip

[CK] Kassel, C.

Quantum Groups, Graduate Texts in Mathematics, Springer

\bigskip

[VC-AP] Chari, V., Pressley, A.

A Guide to Quantum Groups, Cambridge University Press

\bigskip

[SH] Helgason, S.

Differential Geometry, Lie Groups, and Symmetric Spaces

\end{document}